\documentclass[letterpaper,journal]{IEEEtran}
\usepackage{amsmath,amssymb}
\usepackage{graphicx}
\usepackage{booktabs,multirow}
\usepackage{bm}
\usepackage{algorithm}
\usepackage{algpseudocode}
\usepackage[nocompress]{cite}
\usepackage{url}
\usepackage[colorlinks=true,citecolor=blue,linkcolor=black,urlcolor=black]{hyperref}
\newcommand{\ii}{\mathrm{i}}

\begin{document}
\bstctlcite{IEEEcontrol:noURLs}
\setlength{\emergencystretch}{2em}
\title{Conditional Mean Flows for Frequency-Domain Scattered Wavefield Generation}
\author{Lipeng Lai and Shaohuan Zu%
\thanks{Lipeng Lai and Shaohuan Zu are with the Key Laboratory of Earth Exploration and Information Technology of Ministry of Education and the College of Geophysics, Chengdu University of Technology, Chengdu, China.}%
\thanks{Shaohuan Zu is also with the State Key Laboratory of Oil and Gas Reservoir Geology and Exploitation, Chengdu University of Technology, Chengdu, China.}}
\markboth{IEEE Transactions on Geoscience and Remote Sensing}%
{Lai and Zu: Conditional Mean Flows for Frequency-Domain Scattered Wavefield Generation}
\maketitle
\begin{abstract}
Frequency-domain wavefield modeling underpins seismic imaging and inversion, where repeated Helmholtz solutions become costly as velocity models, source locations, and frequencies vary. Neural operators learn mappings between physical fields, and their accuracy can decline outside the training distribution. Diffusion models learn wavefield distributions and generally require repeated sampling evaluations to produce accurate fields. A conditional mean flows model is introduced for complex scattered-wavefield generation. The analytical background wavefield serves as a condition alongside the velocity model, supplying spatially organized information about the source location and frequency. Learning the scattered wavefield avoids the point-source singularity in the target and supports generalization across these physical conditions. The model estimates the average transport velocity over a finite interval connecting Gaussian noise to the scattered field. This representation captures the accumulated displacement and permits both full-interval one-step generation and interval-partitioned sampling. Trained on OpenFWI, the model is evaluated on in-distribution and out-of-distribution velocity structures, frequency generalization, and inference efficiency. In most evaluated settings, one-step mean flows sampling achieves lower wavefield error than multi-step diffusion. Because each sampling step requires another network evaluation, accurate one-step generation also reduces inference time. These results indicate that mean flows can provide accurate and efficient frequency-domain scattered-wavefield generation.
\end{abstract}
\begin{IEEEkeywords}
Frequency-domain modeling, Helmholtz equation, mean flows, neural operators, scattered wavefields.
\end{IEEEkeywords}

\section{Introduction}
\IEEEPARstart{F}{requency-domain} acoustic modeling is a central component of seismic imaging, reverse-time migration, and full-waveform inversion, all of which require wavefields to be evaluated repeatedly as the velocity model, source position, or frequency changes \cite{tarantola1984inversion,virieux2009overview}. By representing the steady-state response at an individual frequency, the frequency-domain formulation is particularly useful for multifrequency and multiscale inversion workflows. Its accuracy is nevertheless critical: in heterogeneous media, small phase errors can accumulate into distorted interference patterns and bias the resulting imaging or inversion. At each frequency, the forward problem reduces to a large, sparse, and indefinite Helmholtz system. Finite-difference, finite-element, and pseudo-spectral schemes provide accurate reference solutions, but their runtime and memory demands grow with model extent, frequency, dimensionality, and the number of sources or right-hand sides \cite{marfurt1984accuracy,virieux1986p,kosloff1982forward,komatitsch1999introduction}. The burden is especially severe for high-resolution three-dimensional models and for workflows that involve many shots and repeated model updates. Although the cost is manageable for an isolated simulation, it becomes a bottleneck in inversion loops, uncertainty studies, and parameter sweeps. The practical challenge is therefore to retain the phase-sensitive accuracy of numerical solvers while making repeated wavefield evaluations substantially cheaper.

Physics-informed neural networks (PINNs) and neural operators have opened two complementary routes toward this goal. PINNs encode the governing equation and boundary conditions in a residual loss, reducing the need for paired labels and providing a mesh-independent coordinate representation \cite{raissi2019physics,karniadakis2021physics}. In seismic modeling, PINNs have been applied to Helmholtz wavefield simulation, time-domain acoustic wave-equation simulation, frequency continuation, solving the frequency-domain acoustic and machine-learned Green's-function solutions \cite{alkhalifah2021wavefield,huang2022pinnup,zou2025accelerating,song2021solving,alkhalifah2020machine}. Their limitation is that solving a new medium or source configuration still requires a costly optimization process, while oscillatory wavefields make spectral bias, residual weighting, and high-frequency convergence difficult to control. Neural operators instead represent mappings between discretized physical fields. In a Fourier neural operator, the integral kernel is parameterized in the Fourier domain, allowing learned spectral interactions to couple distant spatial locations in a field-to-field transformation \cite{li2021fourier}. These methods have been applied to wavefield forward and inverse problems, elastic wave-equation simulation and scattered-wavefield prediction,\cite{yang2021seismic,li2023solving,wang2024transfer,zhang2023learning,zou2024deep,huang2025learned,huang2025pinnfwi}. These models are typically used as deterministic regressors: their accuracy depends on the coverage of the training-data distribution, fine-scale and high-frequency information may be lost during prediction, and generalization to new geological structures, frequencies, or source locations remains difficult.

Generative models offer a probabilistic alternative to deterministic surrogates. Diffusion models first became widely used for high-fidelity image generation by learning the data distribution through a gradual noise-corruption process and reversing that process during sampling \cite{ho2020denoising}. In geophysics, diffusion models have been applied to seismic-data interpolation, seismic-noise attenuation, resolution enhancement, full-waveform-inversion regularization, and velocity-model generation \cite{liu2024geninterp,peng2025fastdiffusion,zhang2024seisresodiff,li2026fwi,feng2024wavediffusion}. For problems governed by partial differential equations (PDEs), this probabilistic formulation is especially useful because it can represent complex solution distributions and condition field reconstruction on physical constraints or partial observations \cite{shu2023physics,huang2024diffusionpde}. More recently, Cheng et al.~\cite{cheng2026seismic} developed a physics-guided generative neural operator for Helmholtz wavefield solutions by incorporating Helmholtz-residual information into conditional diffusion sampling. Diffusion models learn to reverse a sequence of noise levels, so each sample is progressively denoised through multiple Markov transitions and typically requires a network evaluation at each transition \cite{ho2020denoising,song2021ddim}. This repeated sampling increases inference cost, while aggressively reducing the number of steps can degrade wavefield fidelity. These limitations motivate transport formulations with a more direct connection between the noise and data distributions.

Flow Matching addresses this limitation through deterministic transport. It learns a velocity field for an ordinary differential equation (ODE) connecting a Gaussian prior to the data distribution \cite{lipman2023flow,albergo2023stochastic}. Recent flow-matched neural operators extend this idea to learning continuous-time PDE dynamics \cite{hou2026cfo}. Because inference follows this continuous trajectory rather than a long denoising chain, the trajectory can be integrated with larger time steps; the required number of evaluations is determined by the numerical accuracy of the ODE integration and the curvature of the learned marginal trajectory. Thus, Flow Matching can reduce sampling cost, but a coarse solver still introduces integration error. The mean flows model addresses this limitation by learning the displacement accumulated over a finite interval. It parameterizes the interval-average velocity as a function of the current state and the requested interval \cite{geng2025meanflows}. The same parameterization can be evaluated over a sequence of intervals for a controllable rollout, while the full prior-to-data interval gives a one-step operating point. The mean flows identity provides the training relation between average and instantaneous velocities and makes the interval-average field tractable to learn \cite{geng2025meanflows,geng2026improved}.

To reduce the cost of repeated wavefield simulations and generative sampling, a conditional mean flows framework is introduced for complex frequency-domain scattered-wavefield generation \cite{geng2025meanflows,geng2026improved}. The central challenge is to obtain a sufficiently accurate wavefield update without evaluating a generative transport or denoising field at every intermediate step. The mean flows model addresses this challenge by learning the average transport velocity over a finite interval from Gaussian noise to the target field. The interval-average velocity represents the displacement accumulated along the transport path, supporting a full-interval update and partitioned rollouts when additional refinement is required. The condition combines the velocity model with the real and imaginary components of an analytical background wavefield, providing source and frequency information, while the scattered-wavefield target removes the point-source singularity \cite{huang2025learned}. On the OpenFWI benchmark \cite{deng2022openfwi}, the normalized scattered field is connected to Gaussian noise through a conditional linear path. The resulting framework is evaluated on in-distribution and out-of-distribution geological structures, frequency extrapolation, and inference efficiency. The results show that one-step mean flows sampling generally produces more accurate wavefields than multi-step diffusion in the evaluated in-distribution settings and at training frequencies.

\section{Method}

\subsection{The Helmholtz Equation}

Frequency-domain acoustic wave propagation is considered in a two-dimensional heterogeneous medium. Let $U(x,z,\omega)$ denote the complex-valued total wavefield at spatial coordinates $(x,z)$ and angular frequency $\omega$, and let $v(x,z)$ denote the spatially varying physical velocity model. The angular frequency is related to the frequency $f$ in Hz by $\omega=2\pi f$. To separate the source response from the velocity-induced scattering, the total wavefield is decomposed into a background component $U_0$ and a scattered component $\delta U$ \cite{alkhalifah2021wavefield}. The scattered wavefield then satisfies
\begin{equation}
\left\{
\begin{array}{l}
\frac{\omega^2}{v^2}\delta U+\nabla^2\delta U
\,+\omega^2\left(\frac{1}{v^2}-\frac{1}{v_0^2}\right)U_0=0,\\[2pt]
\delta U=U-U_0,
\end{array}
\right.
\label{eq:scattered-helmholtz}
\end{equation}
where $v_0$ is a reference velocity used to construct the background wavefield, $U_0$ is the corresponding solution in the reference medium, and $\nabla^2$ is the spatial Laplacian. This formulation removes the point-source singularity from the learning target and places the velocity dependence in the scattering term.

For a constant reference velocity $v_0$, the background wavefield can be computed analytically from the two-dimensional Green's function:
\begin{equation}
U_0(x,z)=\frac{\ii}{4}H_0^{(2)}\left(\frac{\omega}{v_0}
\sqrt{(x-x_s)^2+(z-z_s)^2}\right),
\label{eq:background-wavefield}
\end{equation}
where $H_0^{(2)}$ denotes the zeroth-order Hankel function of the second kind and $(x_s,z_s)$ is the source location. The analytical background field contains the source-position and frequency information in a spatially organized form. Accordingly, the velocity model and the real and imaginary parts of $U_0$ are used as the input condition, and the real and imaginary parts of $\delta U$ are predicted.

In this study, the velocity model and the real and imaginary components of the analytical background wavefield are used as inputs to predict the corresponding scattered wavefield. This formulation avoids the point-source singularity in the learning target and provides explicit prior information about the source location and frequency, which can improve generalization across source configurations and frequencies \cite{huang2024learned,alkhalifah2021wavefield}. The conditional operator is therefore written as
\begin{equation}
\left(\operatorname{Re}U_0,\operatorname{Im}U_0,v\right)
\longmapsto
\left(\operatorname{Re}\delta U,\operatorname{Im}\delta U\right),
\label{eq:operator-mapping}
\end{equation}
where $\operatorname{Re}(\cdot)$ and $\operatorname{Im}(\cdot)$ denote the real and imaginary parts, respectively. This representation maps physical input fields to frequency-domain scattered responses.

\subsection{Generative Diffusion Model}

Diffusion models define a probabilistic generative process by gradually perturbing data toward a Gaussian prior and learning a reverse process for sample generation \cite{ho2020denoising}. The scattered-wavefield representation provides a target for conditional generative modeling. Let $\mathbf{x}_0$ denote the normalized two-channel scattered wavefield, and define the conditioning field as $\mathbf{C}=\left(\operatorname{Re}U_0,\operatorname{Im}U_0,v\right)$. The model, with trainable parameters $\theta$, approximates the conditional distribution of $\mathbf{x}_0$ given $\mathbf{C}$. The forward process maps each conditional wavefield to a common Gaussian prior, while the reverse process uses $\mathbf{C}$ to transform a Gaussian sample into a scattered wavefield associated with the specified velocity model, source position, and frequency.

In the forward process, Gaussian noise is added through a discrete schedule indexed by $n=1,\ldots,T$, where $T$ is the total number of noise levels. The clean scattered wavefield is denoted by $\mathbf{x}_0$, and $\mathbf{x}_n$ denotes its noisy state at step $n$. Let $\beta_n\in(0,1)$ be the variance increment at step $n$, define $\alpha_n=1-\beta_n$, and let $\bar{\alpha}_n=\prod_{j=1}^{n}\alpha_j$ be the cumulative signal-retention factor. With $\mathbf{I}$ denoting the identity matrix in the field space, one transition of the fixed forward Markov chain is
\begin{equation}
q(\mathbf{x}_n\mid\mathbf{x}_{n-1})
=\mathcal{N}\left(\mathbf{x}_n;\sqrt{\alpha_n}\mathbf{x}_{n-1},\beta_n\mathbf{I}\right).
\label{eq:diffusion-forward}
\end{equation}
Here, $q(\mathbf{x}_n\mid\mathbf{x}_{n-1})$ is the prescribed forward transition, so the next state depends only on the current state and the noise-schedule parameters. Repeated application of these transitions gives the marginal distribution at any noise level:
\begin{equation}
q(\mathbf{x}_n\mid\mathbf{x}_0)
=\mathcal{N}\left(\mathbf{x}_n;\sqrt{\bar{\alpha}_n}\mathbf{x}_0,
 (1-\bar{\alpha}_n)\mathbf{I}\right).
\label{eq:diffusion-closed-form}
\end{equation}
The coefficient $\bar{\alpha}_n$ controls the retained contribution of the clean field, whereas $1-\bar{\alpha}_n$ gives the accumulated noise variance. Equivalently, a noisy field can be sampled using the reparameterization
\begin{equation}
\begin{aligned}
\mathbf{x}_n&=\sqrt{\bar{\alpha}_n}\mathbf{x}_0
+\sqrt{1-\bar{\alpha}_n}\,\boldsymbol{\epsilon},\boldsymbol{\epsilon}\sim\mathcal{N}(\mathbf{0},\mathbf{I}),
\end{aligned}
\label{eq:diffusion-sample}
\end{equation}
where $\boldsymbol{\epsilon}$ is a standard Gaussian field with the same dimensions as $\mathbf{x}_0$. When $\bar{\alpha}_T$ is sufficiently small, $\mathbf{x}_T$ approaches the isotropic Gaussian prior $\mathcal{N}(\mathbf{0},\mathbf{I})$. A training instance therefore consists of a paired field and condition $(\mathbf{x}_0,\mathbf{C})$, a sampled noise level $n$, and the corresponding noise realization $\boldsymbol{\epsilon}$.

The reverse process estimates a clean wavefield from the current noisy state and the physical condition. In the implementation, the network directly predicts $\widehat{\mathbf{x}}_{0,\theta}(\mathbf{x}_n,n\mid\mathbf{C})$. A Gaussian reverse transition can then be written as
\begin{equation}
p_\theta(\mathbf{x}_{n-1}\mid\mathbf{x}_n,\mathbf{C})
=\mathcal{N}\left(\mathbf{x}_{n-1};
\boldsymbol{\mu}_\theta(\mathbf{x}_n,n\mid\mathbf{C}),
\sigma_n^2\mathbf{I}\right),
\label{eq:diffusion-reverse}
\end{equation}
where $p_\theta$ is the learned reverse kernel, $\boldsymbol{\mu}_\theta$ is its mean, and $\sigma_n^2\mathbf{I}$ is its covariance. With $\bar{\alpha}_0=1$, the posterior mean associated with the clean-field prediction is
\begin{equation}
\begin{aligned}
\boldsymbol{\mu}_\theta(\mathbf{x}_n,n\mid\mathbf{C})
&=\frac{\sqrt{\bar{\alpha}_{n-1}}\,\beta_n}{1-\bar{\alpha}_n}
\widehat{\mathbf{x}}_{0,\theta}(\mathbf{x}_n,n\mid\mathbf{C})\\
&\quad+\frac{\sqrt{\alpha_n}(1-\bar{\alpha}_{n-1})}{1-\bar{\alpha}_n}\mathbf{x}_n.
\end{aligned}
\label{eq:diffusion-mean}
\end{equation}
The same forward schedule determines the coefficients in this reverse mean. The direct clean-field prediction is trained with
\begin{equation}
\mathcal{L}_{\mathrm{diff}}
=\mathbb{E}_{\mathbf{x}_0,\mathbf{C},n,\boldsymbol{\epsilon}}
\left[
\left\|
\widehat{\mathbf{x}}_{0,\theta}(\mathbf{x}_n,n\mid\mathbf{C})
-\mathbf{x}_0
\right\|_2^2
\right].
\label{eq:diffusion-objective}
\end{equation}
The objective $\mathcal{L}_{\mathrm{diff}}$ is a denoising regression: it trains the network to recover the clean scattered wavefield $\mathbf{x}_0$ from the noisy state $\mathbf{x}_n$ while using the condition $\mathbf{C}$. Here and below, $\mathbb{E}$ denotes the expectation with respect to the random variables listed in its subscript. For a field vector $\mathbf{q}$ with $N$ scalar entries, $\|\mathbf{q}\|_2=\left(\sum_{i=1}^{N}q_i^2\right)^{1/2}$ denotes the discrete Euclidean norm. At inference, the process starts from $\mathbf{x}_T\sim\mathcal{N}(\mathbf{0},\mathbf{I})$. For the deterministic denoising diffusion implicit model (DDIM) sampler used in the experiments, let $n'$ be the next lower selected noise level and define
\begin{equation}
\widehat{\boldsymbol{\epsilon}}_n
=\frac{\mathbf{x}_n-\sqrt{\bar{\alpha}_n}\,\widehat{\mathbf{x}}_{0,\theta}(\mathbf{x}_n,n\mid\mathbf{C})}
{\sqrt{1-\bar{\alpha}_n}},
\label{eq:diffusion-sampling}
\end{equation}
and update the state according to
\begin{equation}
\begin{aligned}
\mathbf{x}_{n'}&=\sqrt{\bar{\alpha}_{n'}}\,\widehat{\mathbf{x}}_{0,\theta}(\mathbf{x}_n,n\mid\mathbf{C}) + \sqrt{1-\bar{\alpha}_{n'}}\,\widehat{\boldsymbol{\epsilon}}_n.
\end{aligned}
\label{eq:diffusion-ddim-update}
\end{equation}
The final state is taken as the generated clean-field estimate. Because each selected noise level requires a new network evaluation, multi-step DDIM sampling incurs a cost that grows with the number of sampling steps \cite{song2021ddim}. Algorithm~\ref{alg:diffusion} summarizes training and DDIM sampling.

\begin{algorithm}[!t]\small
\caption{Diffusion Training and DDIM Sampling}
\label{alg:diffusion}
\begin{algorithmic}[1]
\Statex \textbf{Training}
\For{each optimization iteration}
  \State Sample $(\mathbf{x}_0,\mathbf{C})$, $n\in\{1,\ldots,T\}$, and $\boldsymbol{\epsilon}\sim\mathcal{N}(\mathbf{0},\mathbf{I})$
  \State $\mathbf{x}_n\gets\sqrt{\bar{\alpha}_n}\mathbf{x}_0+\sqrt{1-\bar{\alpha}_n}\,\boldsymbol{\epsilon}$
  \State Update $\theta$ using $\mathcal{L}_{\mathrm{diff}}$ in Equation~\eqref{eq:diffusion-objective}
\EndFor
\Statex \textbf{DDIM sampling} with $T=n_L>\cdots>n_0=0$
\State Initialize $\mathbf{x}_{n_L}\sim\mathcal{N}(\mathbf{0},\mathbf{I})$
\For{$j=L,\ldots,1$}
  \State Set $n\gets n_j$, $n'\gets n_{j-1}$; predict $\widehat{\mathbf{x}}_{0,\theta}(\mathbf{x}_n,n\mid\mathbf{C})$
  \State Compute $\widehat{\boldsymbol{\epsilon}}_n$ by Equation~\eqref{eq:diffusion-sampling}
  \State Update $\mathbf{x}_{n'}$ by Equation~\eqref{eq:diffusion-ddim-update}
\EndFor
\State \Return $\widehat{\mathbf{x}}_0\gets\mathbf{x}_{n_0}$
\end{algorithmic}
\end{algorithm}

\subsection{Flow Matching}

Diffusion models can be formulated as stochastic differential equations (SDEs), whose reverse dynamics progressively remove noise along potentially complex denoising trajectories. Accurately resolving these dynamics can require many sampling steps. Flow Matching learns a velocity field that defines an ordinary differential equation (ODE) transporting a Gaussian prior to the data distribution. By allowing simple conditional paths, such as linear interpolation between noise and data, this formulation can simplify the transport dynamics and permit larger integration steps, reducing the number of network evaluations required for sampling \cite{lipman2023flow}.

Given the conditioning field $\mathbf{C}$ defined above, let $\boldsymbol{\epsilon}\sim\mathcal{N}(\mathbf{0},\mathbf{I})$ be a Gaussian field and let $\mathbf{x}_0$ denote the normalized two-channel clean scattered wavefield drawn from the conditional data distribution. Here, $\mathbf{x}_0$ corresponds to the clean field in the discrete diffusion formulation, while $\mathbf{z}_0=\boldsymbol{\epsilon}$ denotes the noise endpoint of the continuous transport path. A time variable $t\in[0,1]$ indexes the transport process, where $t=0$ represents the Gaussian noise state and $t=1$ represents the clean scattered wavefield. 

The conditional linear transport path is defined as
\begin{equation}
\mathbf{z}_t=(1-t)\boldsymbol{\epsilon}+t\mathbf{x}_0,
\label{eq:flow-path}
\end{equation}
which connects the Gaussian field at $t=0$ to the clean scattered wavefield
at $t=1$. For fixed endpoints, differentiating with respect to $t$ gives
\begin{equation}
\mathbf{v}^{\mathrm{c}}_t
=\frac{\mathrm{d}\mathbf{z}_t}{\mathrm{d}t}
=\mathbf{x}_0-\boldsymbol{\epsilon}.
\label{eq:flow-conditional-velocity}
\end{equation}
This endpoint difference supplies a constant velocity target for each sampled
pair. A neural velocity field $\mathbf{v}_\theta$, with trainable parameters
$\theta$, predicts the transport velocity from the intermediate field, time,
and physical condition $\mathbf{C}$. It is trained using
\begin{equation}
\mathcal{L}_{\mathrm{FM}}
=\mathbb{E}_{\mathbf{x}_0,\mathbf{C},\boldsymbol{\epsilon},t}
\left\|
\mathbf{v}_{\theta}(\mathbf{z}_t,t\mid\mathbf{C})
-\left(\mathbf{x}_0-\boldsymbol{\epsilon}\right)
\right\|_2^2.
\label{eq:flow-matching-objective}
\end{equation}
Here, $t$ is sampled uniformly from $[0,1]$, and the Gaussian field is sampled
independently of the training wavefield. The target is available directly from
the sampled endpoints, so training requires no trajectory integration. The
learned field approximates the instantaneous transport velocity
$\mathbf{v}(\mathbf{z},t\mid\mathbf{C})$.

During inference, sampling starts from a Gaussian field and follows the
learned velocity field from $t=0$ to $t=1$, with the physical condition
$\mathbf{C}$ held fixed. At each Euler step, the current field is updated by
the predicted velocity multiplied by the step size. The final field gives the
generated scattered wavefield. Each step requires a new network evaluation,
so finer time discretization increases sampling cost.
Algorithm~\ref{alg:flow-matching} summarizes training and Euler sampling.

\begin{algorithm}[!t]\small
\caption{Flow Matching Training and Sampling}
\label{alg:flow-matching}
\begin{algorithmic}[1]

\Statex \textbf{Training}
\For{each optimization iteration}
\State Sample $(\mathbf{x}_0,\mathbf{C})$, 
$\boldsymbol{\epsilon}\sim\mathcal{N}(\mathbf{0},\mathbf{I})$, 
and $t\in[0,1]$
\State $\mathbf{z}_t\gets(1-t)\boldsymbol{\epsilon}+t\mathbf{x}_0$;
$\mathbf{v}^{\mathrm c}\gets\mathbf{x}_0-\boldsymbol{\epsilon}$
\State Update $\theta$ using $\mathcal{L}_{\mathrm{FM}}$ in Equation~\eqref{eq:flow-matching-objective}
\EndFor
\Statex \textbf{Sampling} with $0=t_0<\cdots<t_K=1$
\State Initialize $\mathbf{z}_{t_0}=\boldsymbol{\epsilon}\sim\mathcal{N}(\mathbf{0},\mathbf{I})$
\For{$k=0,\ldots,K-1$}
\State 
$\mathbf{z}_{t_{k+1}}
\gets
\mathbf{z}_{t_k}
+
(t_{k+1}-t_k)
\mathbf{v}_{\theta}(\mathbf{z}_{t_k},t_k\mid\mathbf{C})$
\EndFor
\State \Return $\widehat{\mathbf{x}}_0\gets\mathbf{z}_{t_K}$
\end{algorithmic}
\end{algorithm}

\subsection{Mean Flows}

\begin{figure}[!tb]
\centering
\includegraphics[width=\columnwidth]{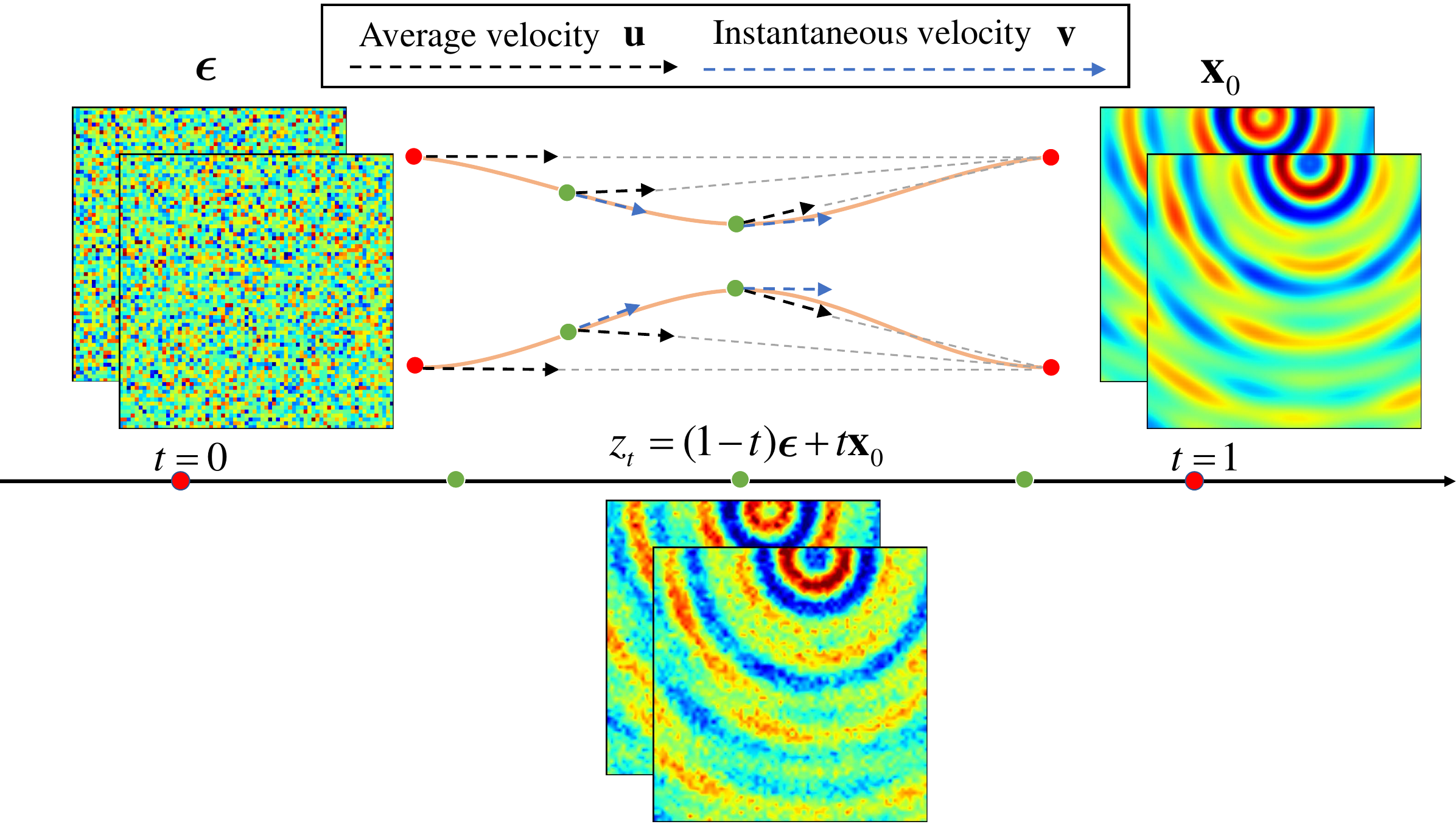}
\caption{Schematic of instantaneous and interval-average transport velocities. The Gaussian state $\mathbf{z}_0=\boldsymbol{\epsilon}$ at $t=0$ is transported to the clean two-channel scattered wavefield $\mathbf{x}_0$ at $t=1$. Blue and black dashed arrows denote local instantaneous velocity $\mathbf{v}$ and finite-interval average velocity $\mathbf{u}$, respectively, with $\mathbf{z}_t-\mathbf{z}_r=(t-r)\mathbf{u}$ over $[r,t]$. Orange curves show marginal ODE trajectories, gray dashed lines show conditional linear paths, and green dots mark intermediate states; the endpoint images show the real and imaginary channels.}
\label{fig:meanflows-schematic}
\end{figure}

Flow Matching learns an instantaneous velocity field along a continuous probability path and generates samples by integrating the corresponding ODE. Because finite displacements must be accumulated through numerical integration, a coarse time discretization can introduce integration error. The mean flows model instead parameterizes the average velocity over a finite interval, as illustrated in Fig.~\ref{fig:meanflows-schematic}. Multiplying this average by the interval length gives the displacement from the Gaussian-noise endpoint $\boldsymbol{\epsilon}$ toward the clean scattered-wavefield endpoint $\mathbf{x}_0$. The instantaneous velocity describes local motion at one time, while the interval-average velocity describes the net transport between two times. Geng et al.~\cite{geng2025meanflows} derive an identity between these velocities, which provides a training relation for the interval-average field. This parameterization supports either a full-interval update or a sequence of shorter updates when a finer numerical approximation is needed.

Consider an interval $[r,t]$ with $0\le r\le t\le1$, where $r$ and $t$ are the start and end times, respectively. For $r<t$, the interval-average velocity is
\begin{equation}
\mathbf{u}(\mathbf{z}_r,r,t\mid\mathbf{C})
=\frac{1}{t-r}\int_r^t
\mathbf{v}(\mathbf{z}_\tau,\tau\mid\mathbf{C})\,\mathrm{d}\tau.
\label{eq:mean-average-velocity}
\end{equation}
For $r=t$, the interval-average velocity is defined by continuity as $\mathbf{u}(\mathbf{z}_r,r,r\mid\mathbf{C})=\mathbf{v}(\mathbf{z}_r,r\mid\mathbf{C})$. The endpoint displacement over $[r,t]$ is therefore
\begin{equation}
(t-r)\mathbf{u}(\mathbf{z}_r,r,t\mid\mathbf{C})
=\int_r^t\mathbf{v}(\mathbf{z}_\tau,\tau\mid\mathbf{C})\,\mathrm{d}\tau.
\label{eq:meanflow-integral-relation}
\end{equation}
Differentiating this relation with respect to the lower endpoint $r$ while holding $t$ and $\mathbf{C}$ fixed gives the mean flows identity
\begin{equation}
\begin{aligned}
\mathbf{v}(\mathbf{z}_r,r\mid\mathbf{C})
&=\mathbf{u}(\mathbf{z}_r,r,t\mid\mathbf{C})\\
&\quad-(t-r)\frac{\mathrm{d}}{\mathrm{d}r}\mathbf{u}(\mathbf{z}_r,r,t\mid\mathbf{C}),
\end{aligned}
\label{eq:meanflow-identity-short}
\end{equation}
where $\mathrm{d}\mathbf{z}_r/\mathrm{d}r=\mathbf{v}(\mathbf{z}_r,r\mid\mathbf{C})$. This form directly relates the instantaneous velocity at the start of the interval to the interval-average velocity and its total derivative. The derivative follows the transport path, so it includes both the change in $\mathbf{z}_r$ and the explicit dependence on $r$. The chain rule gives
\begin{equation}
\frac{\mathrm{d}\mathbf{u}}{\mathrm{d}r}
=\partial_{\mathbf{z}}\mathbf{u}\,\mathbf{v}
+\partial_r\mathbf{u}.
\label{eq:meanflow-total-derivative}
\end{equation}
The state derivative in Equation~\eqref{eq:meanflow-total-derivative} is a Jacobian--vector product (JVP), but it depends on the instantaneous velocity $\mathbf{v}$, which is not directly predicted by the main interval-average head. An auxiliary head $\mathbf{a}_{\theta}(\mathbf{z}_r,r\mid\mathbf{C})$ estimates this velocity while sharing the conditional backbone with $\mathbf{u}_{\theta}$. Using the tangent $(\mathbf{a}_{\theta},1,0)$, which advances the state in the direction $\mathbf{a}_{\theta}$ and increases $r$ at unit rate while holding $t$ fixed, gives
\begin{equation}
\mathbf{J}_\theta
=\partial_{\mathbf{z}}\mathbf{u}_\theta\,\mathbf{a}_\theta
+\partial_r\mathbf{u}_\theta.
\label{eq:meanflow-jvp}
\end{equation}
Substituting this JVP for the total derivative yields the compound instantaneous-velocity prediction
\begin{equation}
\mathbf{V}_\theta
=\mathbf{u}_\theta-(t-r)\operatorname{sg}\left(\mathbf{J}_\theta\right).
\label{eq:meanflow-compound}
\end{equation}
Here $\mathbf{V}_\theta$ combines the interval-average prediction with the finite-interval correction and therefore approximates $\mathbf{v}(\mathbf{z}_r,r\mid\mathbf{C})$. The stop-gradient operator $\operatorname{sg}(\cdot)$ blocks gradients through the JVP correction during optimization but leaves the correction in the forward value. The compound and auxiliary predictions are trained toward the conditional path velocity with the objective
\begin{equation}
\begin{aligned}
\mathcal{L}_{\mathrm{MF}}
&=\mathbb{E}_{\mathbf{x}_0,\mathbf{C},\boldsymbol{\epsilon},r,t}\Bigl[
w\left(\mathbf{V}_\theta-\mathbf{v}^{\mathrm{c}}_r\right)
\left\|\mathbf{V}_\theta-\mathbf{v}^{\mathrm{c}}_r\right\|_2^2\\
&\quad+\lambda_{\mathrm{aux}}
 w\left(\mathbf{a}_\theta-\mathbf{v}^{\mathrm{c}}_r\right)
\left\|\mathbf{a}_\theta-\mathbf{v}^{\mathrm{c}}_r\right\|_2^2\Bigr].
\end{aligned}
\label{eq:meanflow-objective}
\end{equation}
The first term in $\mathcal{L}_{\mathrm{MF}}$ supervises $\mathbf{V}_\theta$, and the second trains $\mathbf{a}_\theta$ for the JVP. The adaptive weight $w(\mathbf{e})=\operatorname{sg}\!\left[(\|\mathbf{e}\|_2^2+\epsilon_{\mathrm{w}})^{-p_{\mathrm{w}}}\right]$ is detached and applied per sample, with $p_{\mathrm{w}}>0$ and $\epsilon_{\mathrm{w}}>0$. The target $\mathbf{v}^{\mathrm{c}}_r=\mathbf{x}_0-\boldsymbol{\epsilon}$ is fixed by the sampled wavefield and Gaussian field. During inference, only the main head is needed because it supplies the interval-average velocity for the transport update. A $K$-step mean flows rollout over $0=t_0<t_1<\cdots t_k \cdots < t_K=1$ is
\begin{equation}
\begin{aligned}
\mathbf{z}_{t_{k+1}}&=\mathbf{z}_{t_k}+(t_{k+1}-t_k)
\mathbf{u}_\theta(\mathbf{z}_{t_k},t_k,t_{k+1}\mid\mathbf{C}).
\end{aligned}
\label{eq:meanflow-multistep}
\end{equation}
The rollout length $K$ is selected according to the desired balance between accuracy and inference cost. The full-interval update is the special case $K=1$: setting $r=0$, $t=1$, and $\mathbf{z}_0=\boldsymbol{\epsilon}$ maps a Gaussian sample directly to the normalized scattered wavefield by
\begin{equation}
\widehat{\mathbf{x}}_0
=\boldsymbol{\epsilon}
+\mathbf{u}_\theta(\boldsymbol{\epsilon},0,1\mid\mathbf{C}),
\label{eq:meanflow-one-step}
\end{equation}
which requires one evaluation of the conditional network and avoids numerical ODE integration in the one-step update. Larger values of $K$ apply the same learned interval-average field over shorter intervals and provide additional transport refinement. Algorithm~\ref{alg:meanflows} summarizes training and sampling. Sampling runs from the Gaussian prior at $t=0$ to the clean wavefield at $t=1$, following the endpoint convention in Fig.~\ref{fig:meanflows-schematic}.

\begin{algorithm}[!t]\small
\caption{Mean Flows Training and Sampling}
\label{alg:meanflows}
\begin{algorithmic}[1]
\Statex \textbf{Training}
\For{each optimization iteration}
  \State Sample $(\mathbf{x_0},\mathbf{C})$, $\boldsymbol{\epsilon}\sim\mathcal{N}(\mathbf{0},\mathbf{I})$, and $0\le r\le t\le1$
  \State $\mathbf{z}_r\gets(1-r)\boldsymbol{\epsilon}+r\mathbf{x_0}$; $\mathbf{v}^{\mathrm c}\gets\mathbf{x_0}-\boldsymbol{\epsilon}$
  \State Evaluate $\mathbf{u}_\theta(\mathbf{z}_r,r,t\mid\mathbf{C})$ and $\mathbf{a}_\theta(\mathbf{z}_r,r\mid\mathbf{C})$
  \State $\mathbf{J}_\theta\gets\operatorname{JVP}_{(\mathbf{z},r,t)}\!\left[\mathbf{u}_\theta;(\mathbf{a}_\theta,1,0)\right]$
  \State $\mathbf{V}_\theta\gets\mathbf{u}_\theta-(t-r)\operatorname{sg}(\mathbf{J}_\theta)$
  \State Update $\theta$ using $\mathcal{L}_{\mathrm{MF}}$ in Equation~\eqref{eq:meanflow-objective}
\EndFor
\Statex \textbf{Sampling} with $0=t_0<\cdots<t_K=1$
\State Initialize $\mathbf{z}_{t_0}=\boldsymbol{\epsilon}\sim\mathcal{N}(\mathbf{0},\mathbf{I})$
\For{$k=0,\ldots,K-1$}
  \State $\mathbf{z}_{t_{k+1}}\gets\mathbf{z}_{t_k}+(t_{k+1}-t_k)\mathbf{u}_\theta(\mathbf{z}_{t_k},t_k,t_{k+1}\mid\mathbf{C})$
\EndFor
\State \Return $\widehat{\mathbf{x_0}}\gets\mathbf{z}_{t_K}$
\end{algorithmic}
\end{algorithm}

\section{Results}

\subsection{Evaluation Metrics}

The relative $L_2$ error (Rel $L_2$) is used as the primary evaluation metric. For a predicted wavefield $\widehat{\delta U}$ and the corresponding reference wavefield ${\delta U}$, it is defined as
\begin{equation}
\mathrm{Rel} \mathit{L}_{2} =\frac{\left|\left|\widehat{ \delta U}- \delta U\right|\right|_{ 2 }}{\left|\left| \delta U\right|\right|_{ 2 }},
\label{eq:relative-l2}
\end{equation}
where $\|\cdot\|_2$ is the discrete Euclidean norm defined above. Rel $L_2$ is dimensionless and measures the prediction error relative to the reference magnitude, with smaller values indicating more accurate predictions.

To quantify inference efficiency, the speed-up ratio is defined as
\begin{equation}
S=\frac{T_{\mathrm{ref}}}{T_{\mathrm{method}}},
\label{eq:speedup}
\end{equation}
where $T_{\mathrm{ref}}$ and $T_{\mathrm{method}}$ denote the runtimes of the reference numerical simulation and the evaluated generative model, respectively. A value of $S>1$ indicates acceleration relative to the reference.

\subsection{Experimental Configuration}

The dataset for frequency-domain acoustic wave propagation is constructed from the CurveVel-A and FlatFault-A families of OpenFWI, using the four subsets CurveVel-A-1, CurveVel-A-2, FlatFault-A-1, and FlatFault-A-2 \cite{deng2022openfwi}. Each subset contains 500 original velocity models, of which 480 are assigned to training and 20 to validation before augmentation. Each original model is paired with one Gaussian-smoothed variant, with the smoothing standard deviation sampled uniformly between 1 and 2 grid cells. Original models and their smoothed variants remain in the same split, giving 3,840 training models and 160 validation models. Each model is discretized on a $70\times70$ grid with $0.01$~km spacing and paired with all integer frequencies from 3 to 16~Hz and four fixed source locations at grid indices $(z,x)=(2,21)$, $(2,31)$, $(2,41)$, and $(2,51)$. An analytical Green's function for a constant background velocity of $1.5$~km/s supplies the real and imaginary background-wavefield channels. For every model--frequency--source combination, the scattered-field Helmholtz equation is solved with a nine-point isotropic finite-difference stencil \cite{jo1996}, a 12-cell absorbing boundary layer, and a sparse direct solver. The resulting two-channel scattered fields yield 215,040 training samples and 8,960 validation samples.

The diffusion and mean flows models share the same conditional U-Net backbone. The network takes two state channels together with three conditioning channels---the real and imaginary background fields and the velocity model---and produces a two-channel scattered-wavefield prediction. The base width is 29, the channel multipliers are $(1,2,4,8,16)$, and each resolution level contains two residual blocks. Attention uses four heads at downsampling factors of 4, 8, and 16. The diffusion model uses a 1,000-step cosine noise schedule and directly predicts $\mathbf{x}_0$, followed by deterministic DDIM sampling with $\eta=0$. The mean flows model uses the same spatial backbone, augmented with two time inputs and an auxiliary instantaneous-velocity head. Its improved training formulation adopts a linear transport path and logit-normal endpoint sampling with $(\mu_{\mathrm{LN}},\sigma_{\mathrm{LN}})=(-0.4,1.0)$. Unequal endpoints are used for 50\% of samples, including the full interval $(r,t)=(0,1)$ for 15\% of all samples; the remaining samples use $r=t$. Adaptive weighting uses $(p_{\mathrm{w}},\epsilon_{\mathrm{w}})=(1.0,0.01)$, with unit auxiliary-velocity loss weight and a time-input scale of 1000. A deterministic Fourier neural operator (FNO) \cite{li2021fourier} provides a direct-regression baseline. It has four Fourier layers of width 128, retains 24 modes along each spatial dimension, uses rank-22 spectral factors, and includes coordinate channels and eight-cell padding. FNO has approximately 52.3 million parameters, compared with approximately 51.6 million for each generative model, and produces a prediction in one network evaluation.

\begin{figure}[!tb]
\centering
\includegraphics[width=\columnwidth]{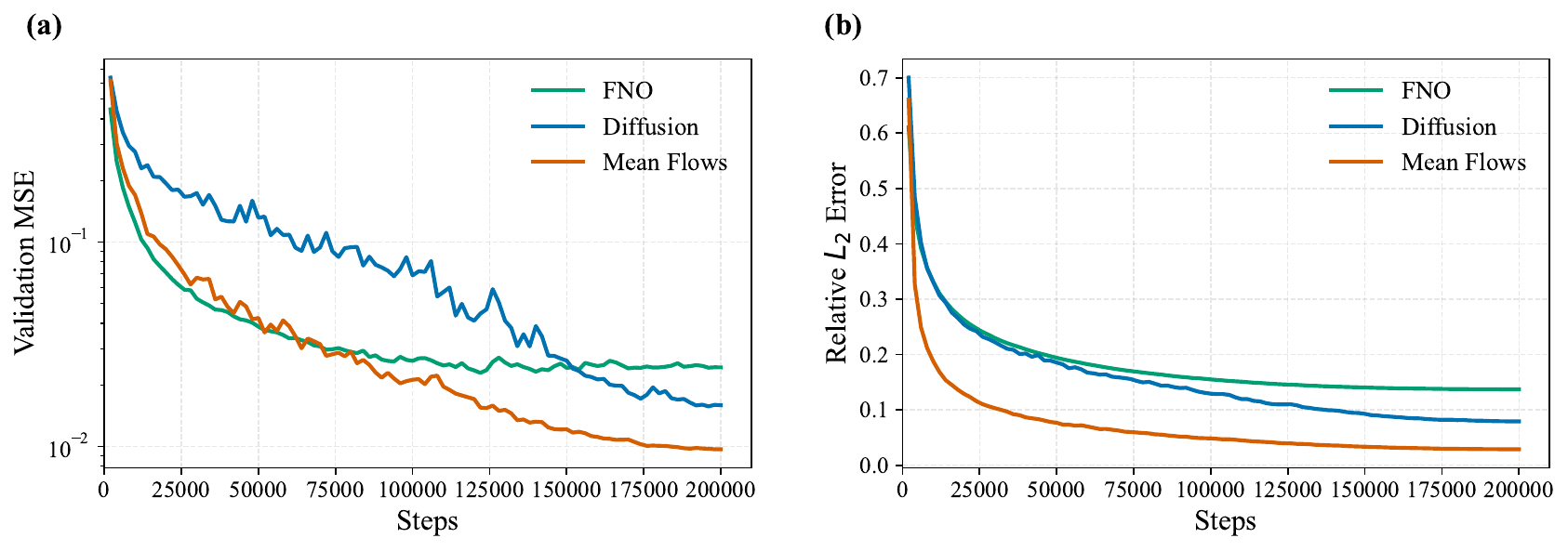}
\caption{Validation histories of FNO, diffusion, and mean flows over 200,000 training steps. Panel (a) shows standardized-target MSE on a logarithmic axis; panel (b) shows mean per-sample relative $L_2$ error in the original wavefield scale on a linear axis.}
\label{fig:loss-curves}
\end{figure}

All three models are trained for 200,000 optimization steps with an initial learning rate of $10^{-4}$. Diffusion and mean flows use batch size 16 and AdamW weight decay $5\times10^{-5}$; FNO uses batch size 8 and weight decay $10^{-4}$. All methods use $(\beta_1,\beta_2)=(0.9,0.999)$, cosine learning-rate annealing to $10^{-6}$, and an exponential moving average with decay 0.999 for evaluation. Fig.~\ref{fig:loss-curves} compares validation mean squared error (MSE) in the standardized target space on a logarithmic axis with relative $L_2$ error on a linear axis. These metrics weight sample amplitudes differently and therefore need not exhibit identical convergence profiles. In particular, FNO has a spectral bias toward low-frequency components: early in training, it can reproduce the dominant low-frequency content of the scattered wavefield relatively well, which may yield a comparatively small MSE. As training proceeds, however, its limited recovery of high-frequency and fine-scale components restricts further improvement, allowing diffusion and mean flows to overtake it. The relative $L_2$ error is normalized by the $L_2$ norm of the reference field for each sample, so it measures proportional whole-field error and is less sensitive to differences in absolute amplitude across samples; its convergence can therefore differ from that of the amplitude-weighted MSE. At 200,000 steps, the normalized MSE and relative $L_2$ error are 0.00969 and 0.02883 for mean flows, 0.01590 and 0.07928 for diffusion, and 0.02435 and 0.13713 for FNO. The spatial and frequency accuracy comparisons below use 1, 2, 4, and 8 sampling steps for the generative models; FNO is evaluated only as a single-pass predictor. Diffusion and mean flows use the same initial noise for each paired comparison.

\subsection{In-Distribution Test}

The in-distribution evaluation tests wavefield generation on held-out velocity models from the same geological families as the training data. CurveVel-A and FlatFault-A cases are evaluated at 12 and 14~Hz, respectively. CurveVel-A has smoothly varying, curved interfaces and gradual lateral velocity changes, whereas FlatFault-A combines approximately horizontal layers with fault-induced discontinuities and sharper local velocity contrasts. These cases assess prediction accuracy on both smoothly varying and faulted structures represented in the training distribution.

\begin{figure}[!tb]
\centering
\includegraphics[width=\columnwidth]{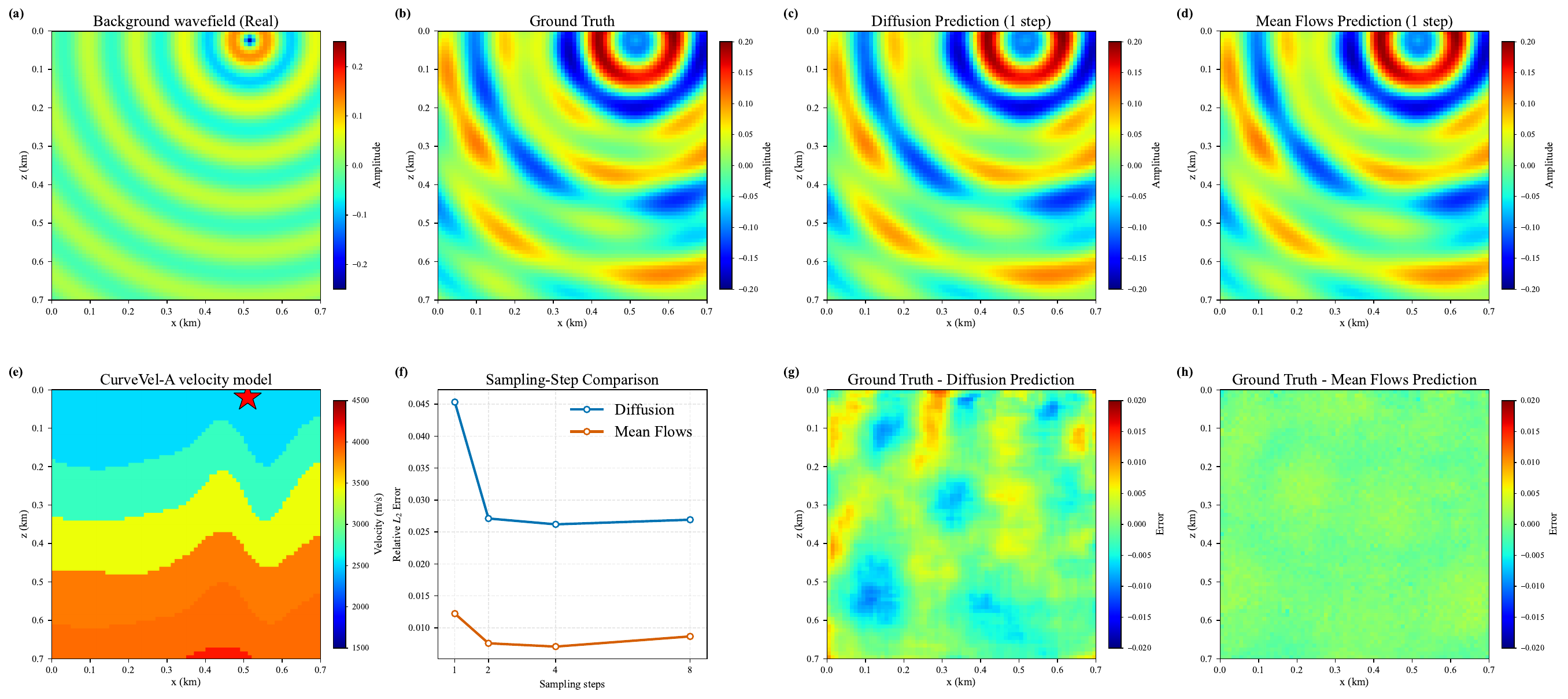}
\caption{In-distribution scattered-wavefield generation on CurveVel-A at 12~Hz. Panels (a)--(e) show the real background field, reference scattered field, one-step diffusion prediction, one-step mean flows prediction, and velocity model, respectively. Panel (f) reports complex-field relative $L_2$ errors at 1, 2, 4, and 8 sampling steps. Panels (g) and (h) show the real-component reference-minus-prediction fields for diffusion and mean flows.}
\label{fig:curvevel-a-comparison}
\end{figure}

\begin{figure}[!tb]
\centering
\includegraphics[width=\columnwidth]{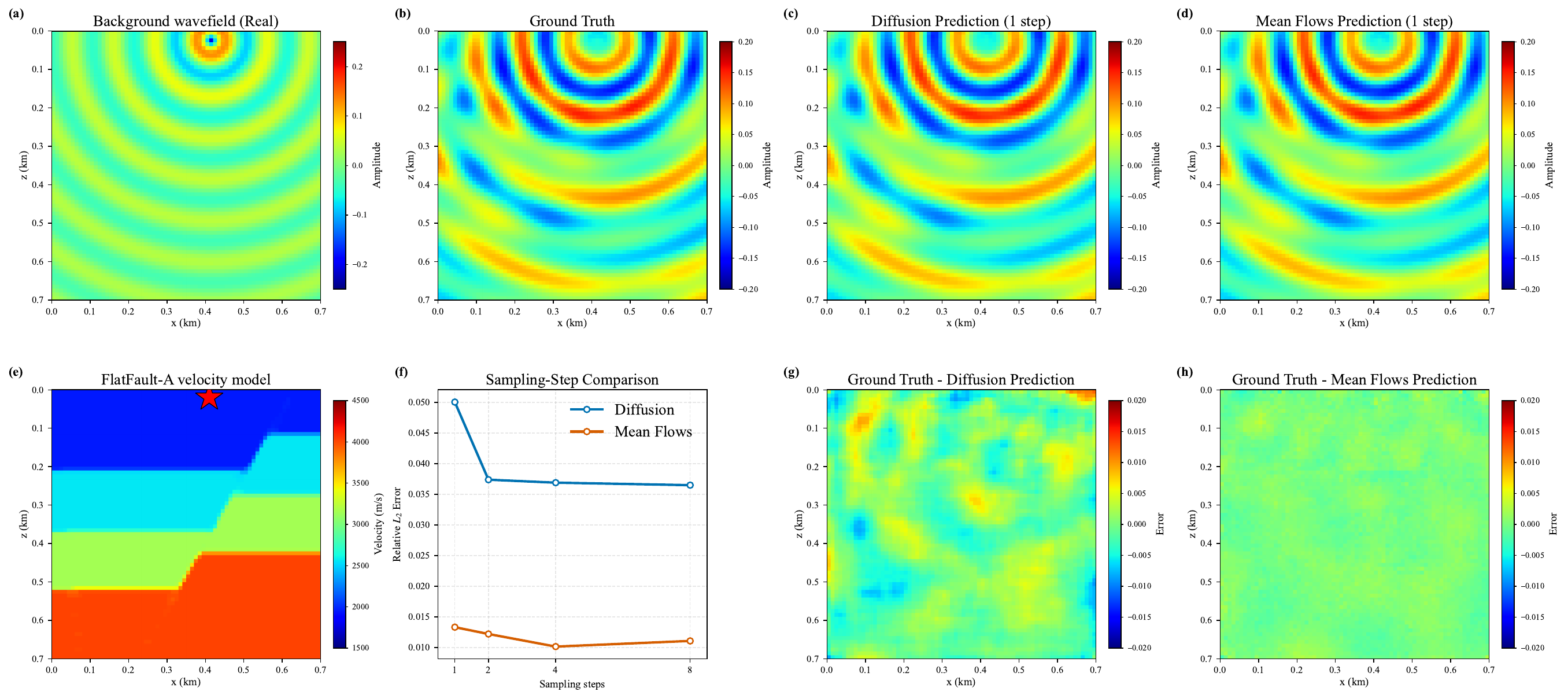}
\caption{In-distribution scattered-wavefield generation on FlatFault-A at 14~Hz. Panels (a)--(e) show the real background field, reference scattered field, one-step diffusion prediction, one-step mean flows prediction, and velocity model, respectively. Panel (f) reports complex-field relative $L_2$ errors at 1, 2, 4, and 8 sampling steps. Panels (g) and (h) show the real-component reference-minus-prediction fields for diffusion and mean flows.}
\label{fig:flatfault-a-comparison}
\end{figure}

Figs.~\ref{fig:curvevel-a-comparison} and~\ref{fig:flatfault-a-comparison} compare the predicted scattered wavefields and their residuals. Panel (a) shows the real analytical background field, panel (b) the Helmholtz reference solution, panels (c) and (d) the one-step diffusion and mean flows predictions, panel (e) the velocity model, panel (f) the relative $L_2$ error versus sampling steps, and panels (g) and (h) the reference-minus-prediction fields. The residual color range is one-tenth of the prediction color range, making local discrepancies more visible. One-step mean flows gives relative $L_2$ errors of 0.0122 for CurveVel-A and 0.0133 for FlatFault-A, compared with 0.0453 and 0.0501 for one-step diffusion. These mean flows errors also remain below the corresponding eight-step diffusion errors of 0.0269 and 0.0365. The displayed real-component residuals are consistent with these complex-field metrics, with weaker coherent phase and amplitude discrepancies for mean flows.

\begin{figure}[!tb]
\centering
\includegraphics[width=\columnwidth]{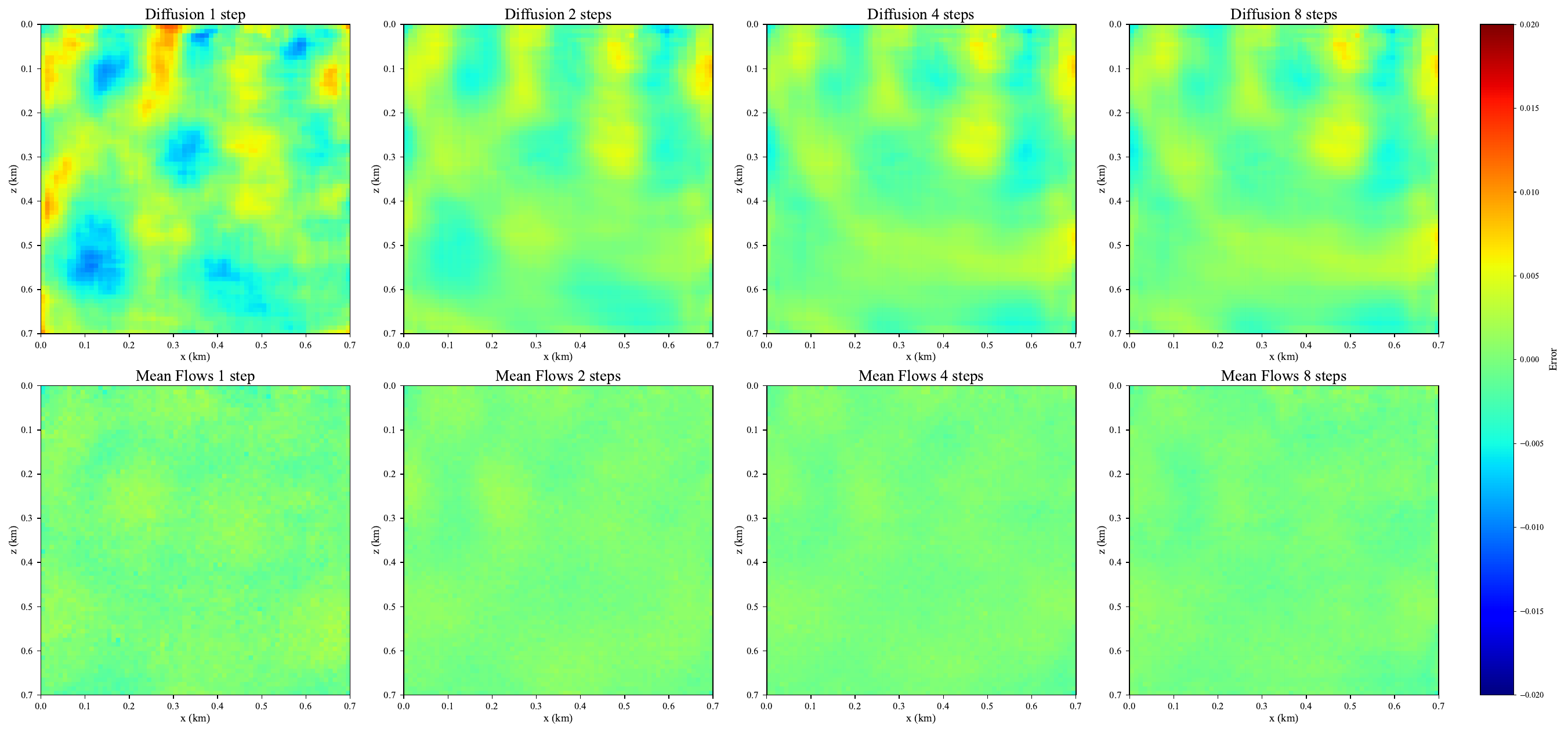}
\caption{Sampling-step residual comparison on CurveVel-A at 12~Hz for the same model and source as Fig.~\ref{fig:curvevel-a-comparison}. The upper row shows diffusion and the lower row shows mean flows at 1, 2, 4, and 8 steps from left to right.}
\label{fig:curvevel-a-comparison-step}
\end{figure}

\begin{figure}[!tb]
\centering
\includegraphics[width=\columnwidth]{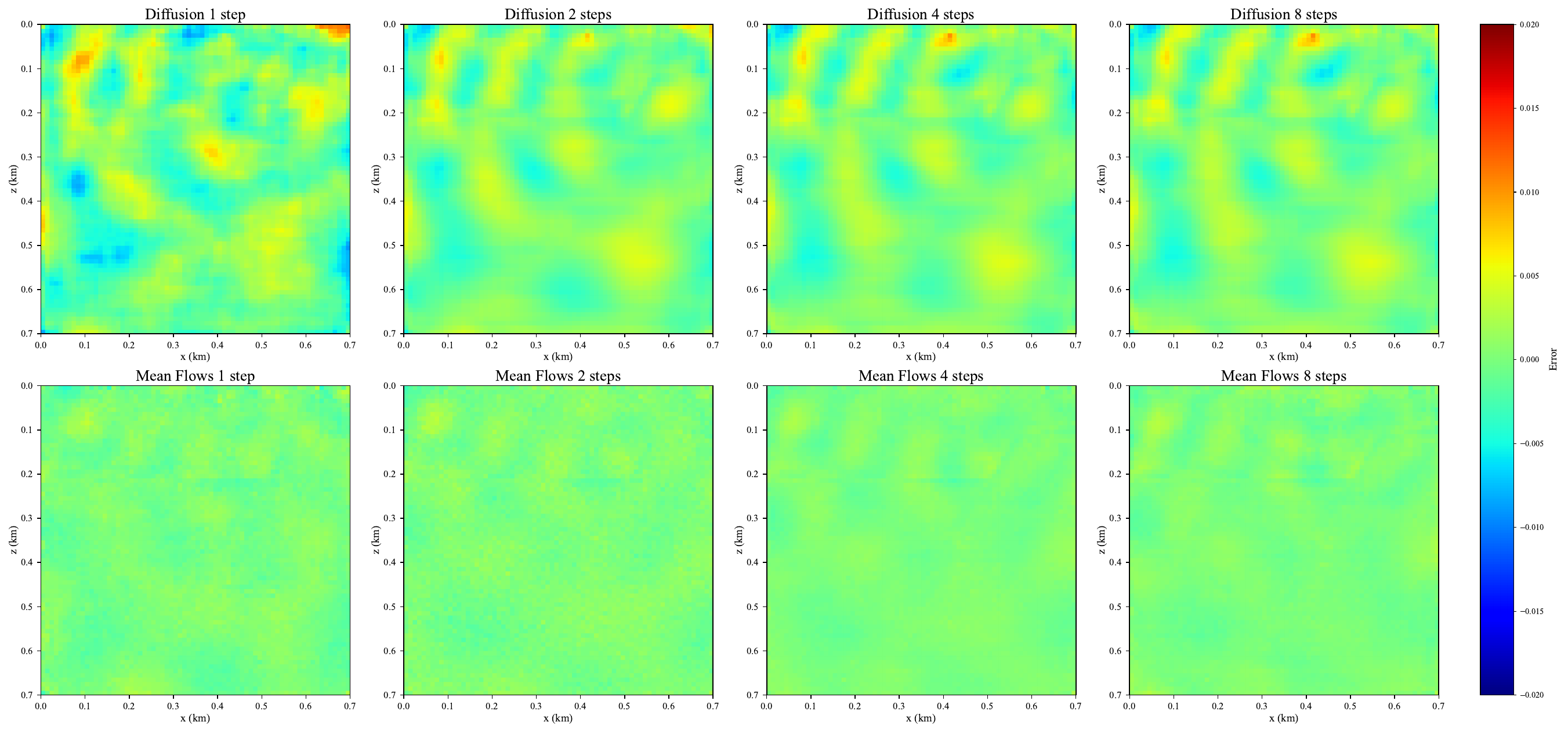}
\caption{Sampling-step residual comparison on FlatFault-A at 14~Hz for the same model and source as Fig.~\ref{fig:flatfault-a-comparison}. The upper row shows diffusion and the lower row shows mean flows at 1, 2, 4, and 8 steps from left to right.}
\label{fig:flatfault-a-comparison-step}
\end{figure}

Figs.~\ref{fig:curvevel-a-comparison-step} and~\ref{fig:flatfault-a-comparison-step} show the residuals at 1, 2, 4, and 8 sampling steps, with diffusion in the upper rows and mean flows in the lower rows. Additional diffusion updates reduce the dominant error structures, although coherent residuals remain at eight steps. For mean flows, partitioning the transport interval improves on the one-step prediction, with the lowest errors in these two cases occurring at four steps. The small increase from four to eight steps shows that a longer rollout does not guarantee a more accurate prediction. The main advantage in these examples is that a single mean flows update already gives lower error than every tested diffusion rollout.

\subsection{Out-of-Distribution Test}

The out-of-distribution evaluation examines transfer to geological structures absent from the training families. The Marmousi and Overthrust models are evaluated at 16~Hz, which lies within the training-frequency range so that the test isolates geological transfer. Marmousi contains strongly undulating and laterally varying reflectors with localized velocity anomalies, producing complex interference and diffraction patterns. Overthrust contains thrust-related interfaces, steeply dipping structures, and strong velocity contrasts. Their spatial organization differs from the CurveVel-A and FlatFault-A structures used for training.

\begin{figure}[!tb]
\centering
\includegraphics[width=\columnwidth]{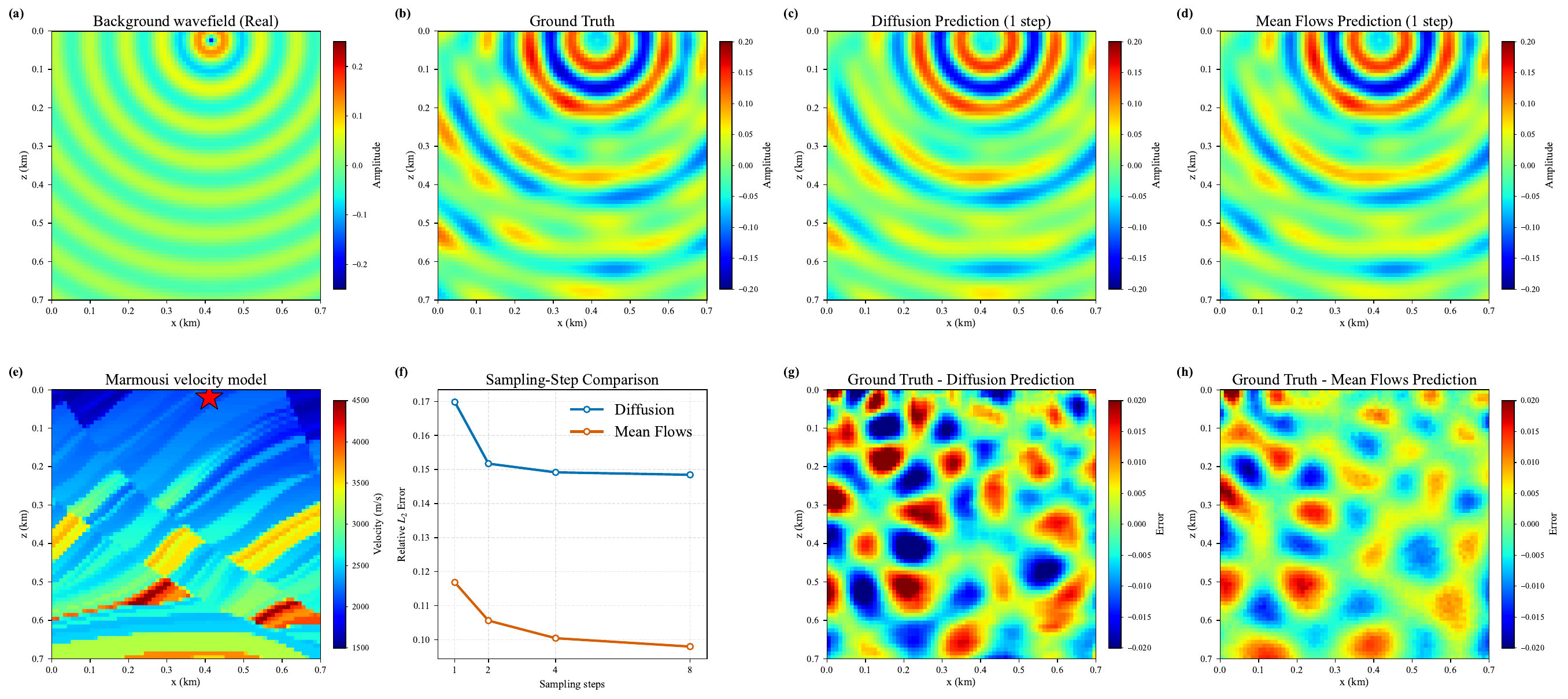}
\caption{Out-of-distribution scattered-wavefield generation on Marmousi at 16~Hz. Panels (a)--(e) show the real background field, reference scattered field, one-step diffusion prediction, one-step mean flows prediction, and velocity model, respectively. Panel (f) reports complex-field relative $L_2$ errors at 1, 2, 4, and 8 sampling steps. Panels (g) and (h) show the real-component reference-minus-prediction fields for diffusion and mean flows.}
\label{fig:marmousi-comparison}
\end{figure}

\begin{figure}[!tb]
\centering
\includegraphics[width=\columnwidth]{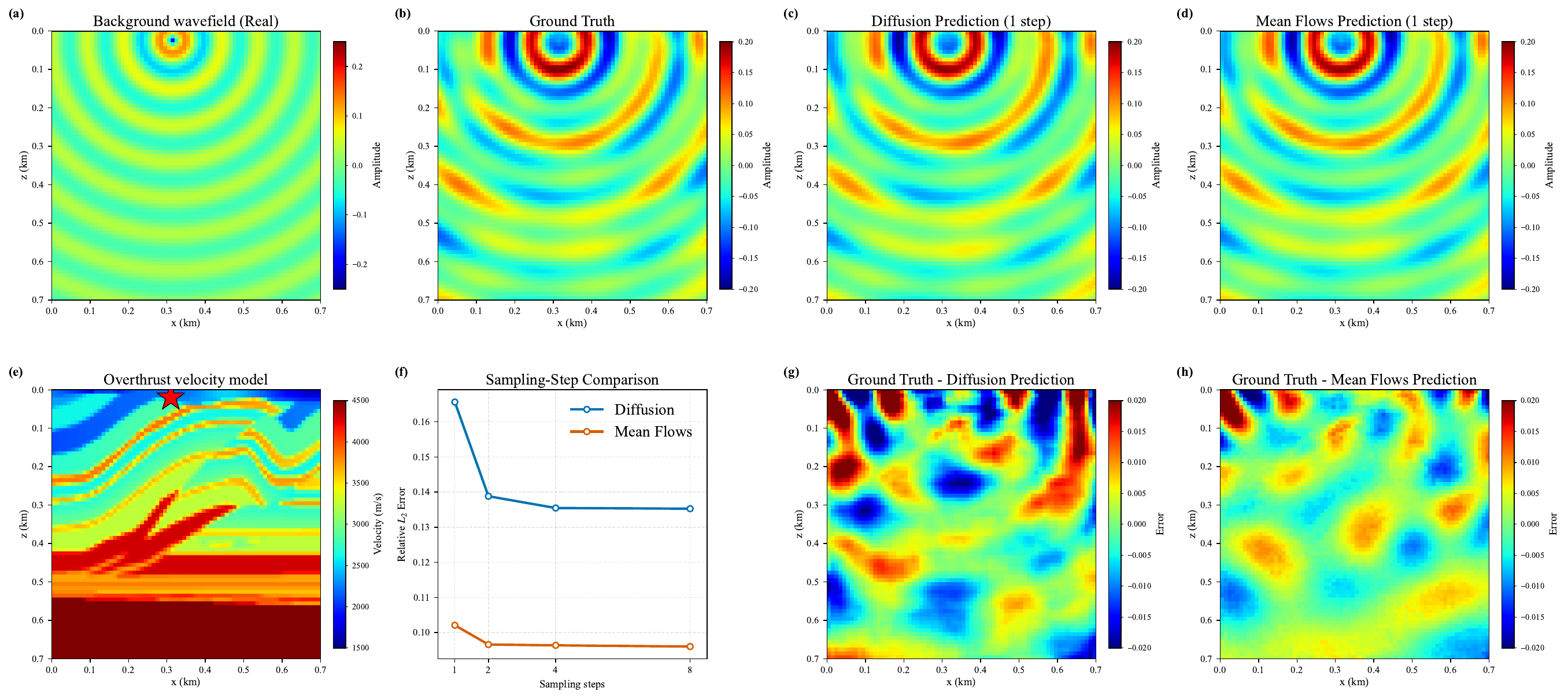}
\caption{Out-of-distribution scattered-wavefield generation on Overthrust at 16~Hz. Panels (a)--(e) show the real background field, reference scattered field, one-step diffusion prediction, one-step mean flows prediction, and velocity model, respectively. Panel (f) reports complex-field relative $L_2$ errors at 1, 2, 4, and 8 sampling steps. Panels (g) and (h) show the real-component reference-minus-prediction fields for diffusion and mean flows.}
\label{fig:overthrust-comparison}
\end{figure}

Figs.~\ref{fig:marmousi-comparison} and~\ref{fig:overthrust-comparison} use the same panel organization as the in-distribution comparisons, including one-step predictions in panels (c) and (d), sampling-step error curves in panel (f), and reference-minus-prediction fields in panels (g) and (h). On Marmousi, the one-step relative $L_2$ errors are 0.1168 for mean flows and 0.1698 for diffusion; on Overthrust, they are 0.1021 and 0.1656, respectively. Mean flows also gives weaker residual structures in the displayed real-component fields. Both methods have larger errors than in the selected in-distribution examples, indicating that transfer to unfamiliar geology remains more difficult even at a training frequency.

\begin{figure}[!tb]
\centering
\includegraphics[width=\columnwidth]{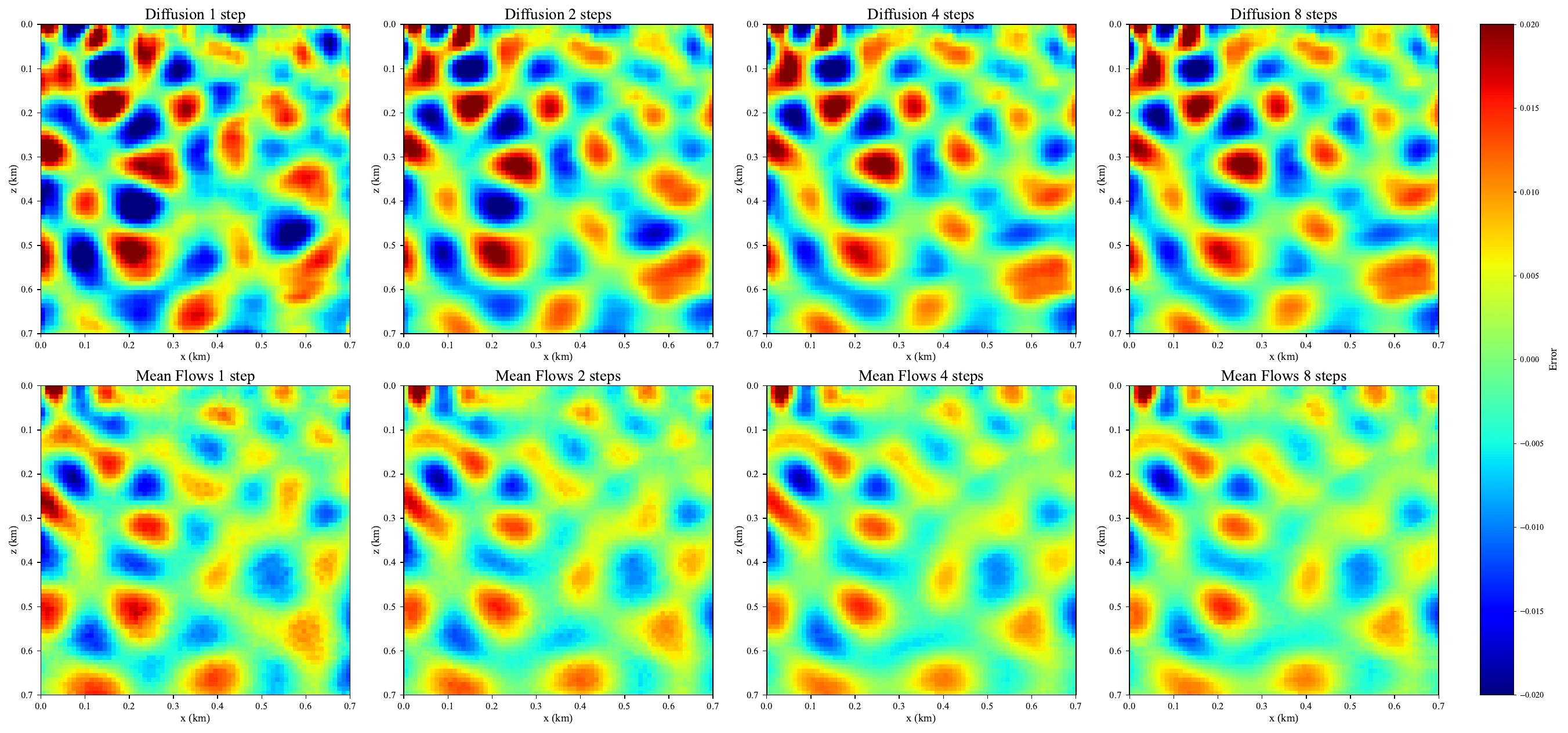}
\caption{Sampling-step residual comparison on Marmousi at 16~Hz for the same model and source as Fig.~\ref{fig:marmousi-comparison}. The upper row shows diffusion and the lower row shows mean flows at 1, 2, 4, and 8 steps from left to right.}
\label{fig:marmousi-comparison-step}
\end{figure}

\begin{figure}[!tb]
\centering
\includegraphics[width=\columnwidth]{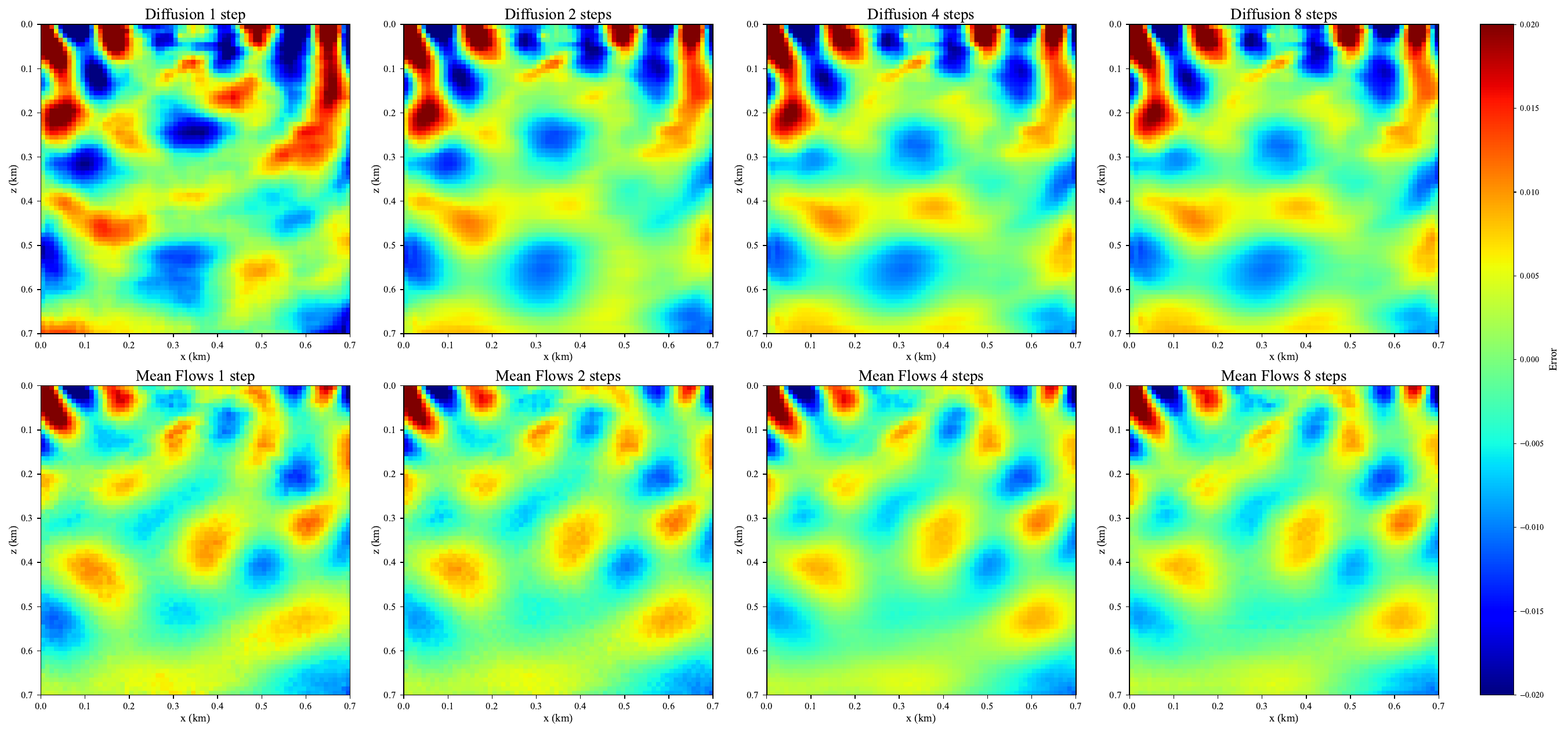}
\caption{Sampling-step residual comparison on Overthrust at 16~Hz for the same model and source as Fig.~\ref{fig:overthrust-comparison}. The upper row shows diffusion and the lower row shows mean flows at 1, 2, 4, and 8 steps from left to right.}
\label{fig:overthrust-comparison-step}
\end{figure}

The residual sequences in Figs.~\ref{fig:marmousi-comparison-step} and~\ref{fig:overthrust-comparison-step} show the effect of 1, 2, 4, and 8 sampling steps. Additional steps improve both methods in these cases, but one-step mean flows remains more accurate than eight-step diffusion: 0.1168 versus 0.1485 on Marmousi and 0.1021 versus 0.1353 on Overthrust. Eight-step mean flows further reduces the errors to 0.0980 and 0.0960, respectively. The smaller improvement on Overthrust suggests that most of its mean flows accuracy is already obtained from the full-interval update, while Marmousi benefits more from interval partitioning. 

\subsection{Frequency Extrapolation Test}

\begin{figure}[!tb]
\centering
\includegraphics[width=\columnwidth]{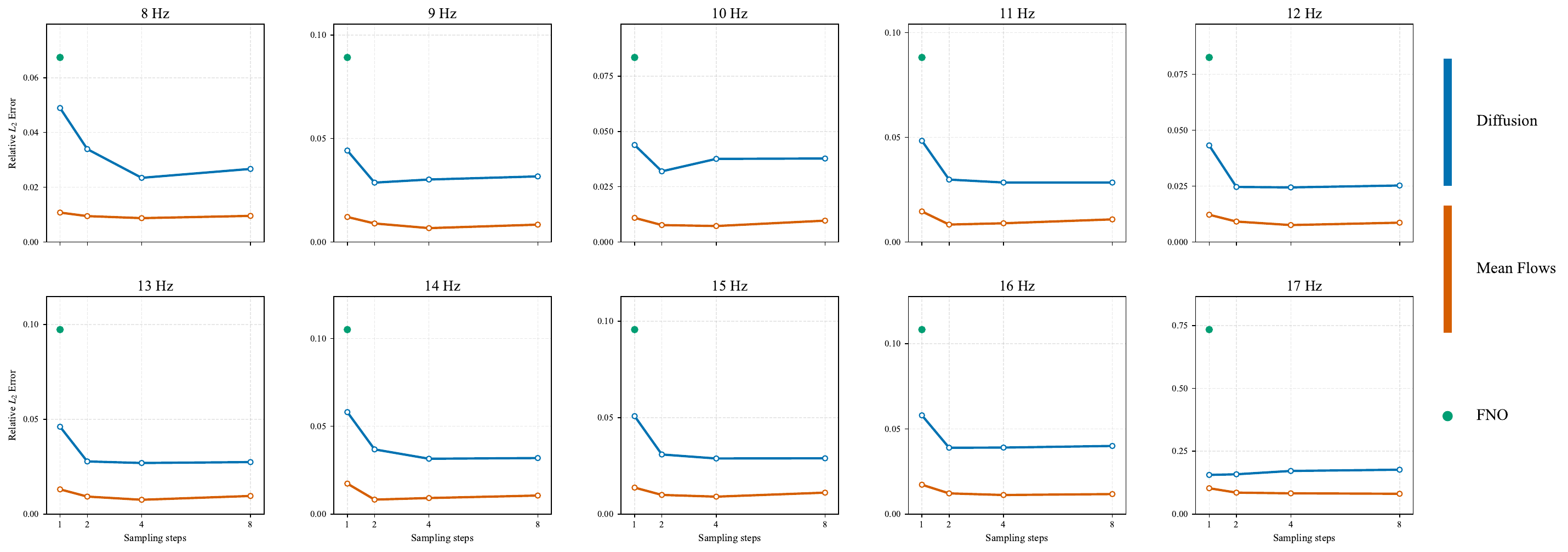}
\caption{Frequency-dependent sampling accuracy on CurveVel-A. The upper row shows 8--12~Hz and the lower row 13--17~Hz, ordered from left to right. Frequencies 8--16~Hz are represented in training. Blue and orange curves show relative $L_2$ errors for diffusion and mean flows at 1, 2, 4, and 8 steps; the green point at one step shows the single-pass FNO prediction.}
\label{fig:sampling-curves-curvevel-a}
\end{figure}

\begin{figure}[!tb]
\centering
\includegraphics[width=\columnwidth]{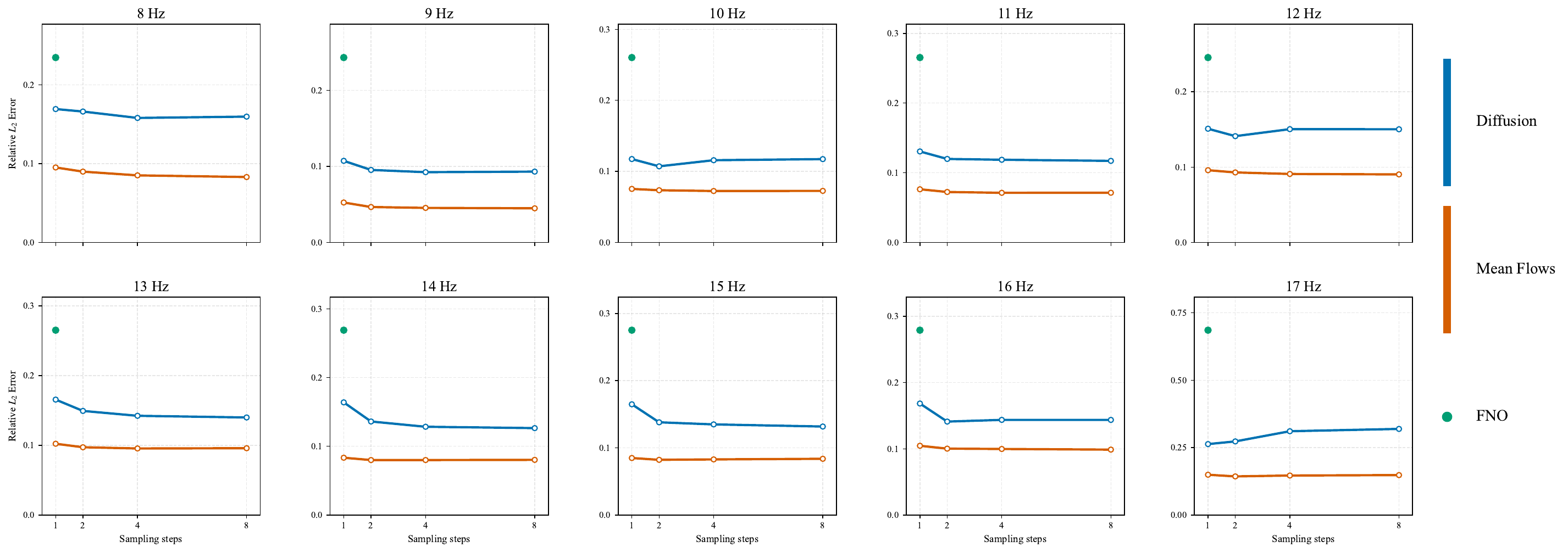}
\caption{Frequency-dependent sampling accuracy on Overthrust. The upper row shows 8--12~Hz and the lower row 13--17~Hz, ordered from left to right. Frequencies 8--16~Hz are represented in training. Blue and orange curves show relative $L_2$ errors for diffusion and mean flows at 1, 2, 4, and 8 steps; the green point at one step shows the single-pass FNO prediction.}
\label{fig:sampling-curves-overthrust}
\end{figure}

Figs.~\ref{fig:sampling-curves-curvevel-a} and~\ref{fig:sampling-curves-overthrust} compare frequency-dependent accuracy on a held-out CurveVel-A model and the unseen Overthrust model, respectively. Each figure covers 8--17~Hz, with 8--12~Hz arranged from left to right in the upper row and 13--17~Hz in the lower row. Because training includes every integer frequency from 3 to 16~Hz, the 8--16~Hz panels evaluate frequencies represented during training, while 17~Hz evaluates extrapolation. Diffusion and mean flows are compared at 1, 2, 4, and 8 sampling steps; FNO is shown only by a green point at one network evaluation. Mean flows has lower one-step error than FNO and every tested diffusion rollout at each displayed frequency for both selected models. At 17~Hz, the one-step errors for mean flows, diffusion, and FNO are 0.1028, 0.1557, and 0.7342 on CurveVel-A, and 0.1492, 0.2629, and 0.6853 on Overthrust. The increase relative to the neighboring 16~Hz results nevertheless shows a loss of accuracy beyond the training range. Additional diffusion steps do not remove this increase and can raise the error, while mean flows retains a smaller error with modest changes across rollout lengths.

\begin{figure}[!tb]
\centering
\includegraphics[width=\columnwidth]{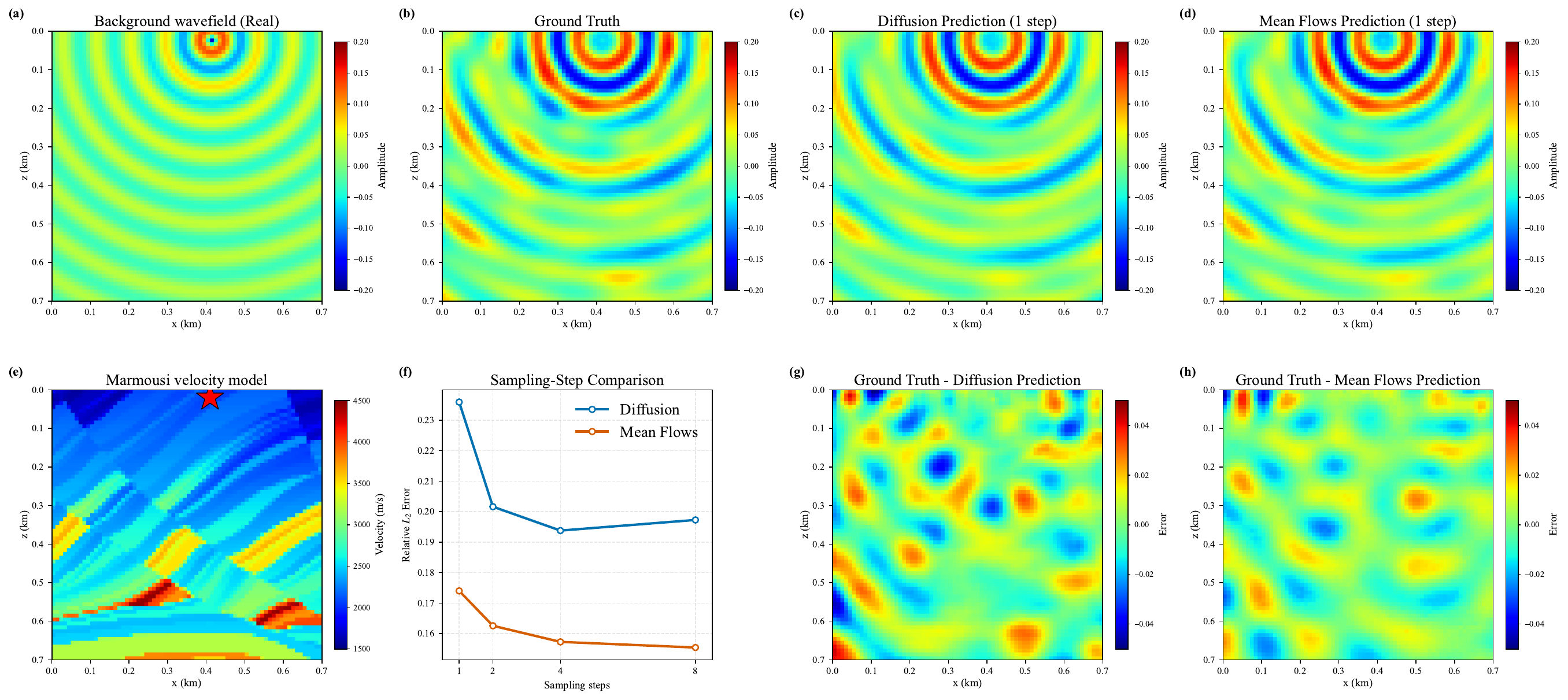}
\caption{Frequency extrapolation on the unseen Marmousi model at 17~Hz. Panels (a)--(e) show the real background field, reference scattered field, one-step diffusion prediction, one-step mean flows prediction, and velocity model, respectively. Panel (f) reports relative $L_2$ errors at 1, 2, 4, and 8 steps; panels (g) and (h) show the one-step real-component reference-minus-prediction fields.}
\label{fig:marmousi-17hz-comparison}
\end{figure}

\begin{figure}[!tb]
\centering
\includegraphics[width=\columnwidth]{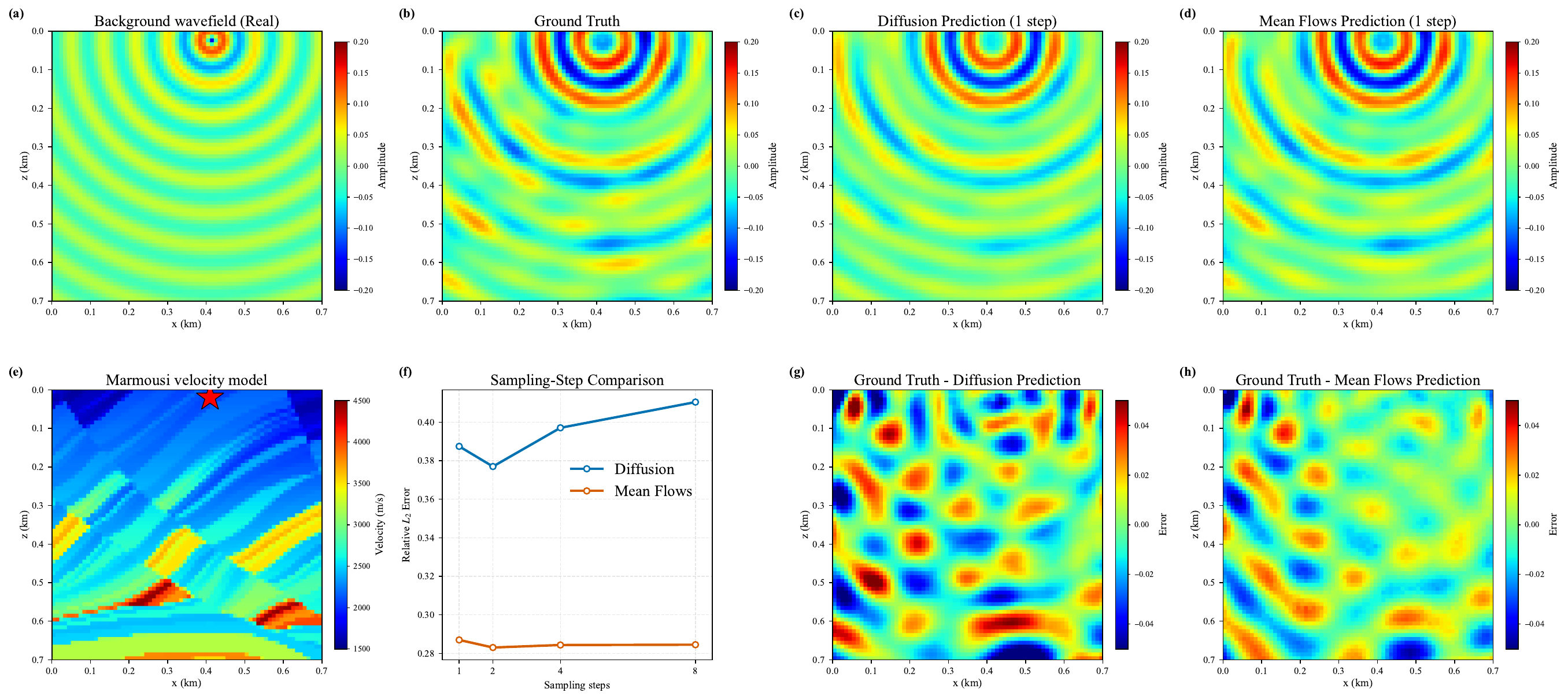}
\caption{Frequency extrapolation on the unseen Marmousi model at 18~Hz. Panels (a)--(e) show the real background field, reference scattered field, one-step diffusion prediction, one-step mean flows prediction, and velocity model, respectively. Panel (f) reports relative $L_2$ errors at 1, 2, 4, and 8 steps; panels (g) and (h) show the one-step real-component reference-minus-prediction fields.}
\label{fig:marmousi-18hz-comparison}
\end{figure}

Figs.~\ref{fig:marmousi-17hz-comparison} and~\ref{fig:marmousi-18hz-comparison} examine the spatial predictions at 17 and 18~Hz on Marmousi, combining an unseen geological model with frequencies outside the training range. Both use the same source and panel arrangement as the 16~Hz Marmousi comparison. At 17~Hz, one-step mean flows gives a relative $L_2$ error of 0.1740, compared with 0.2359 for diffusion; at 18~Hz, the errors rise to 0.2870 and 0.3875, respectively. The wavefields retain their main oscillatory structure, but the residual panels show increasing local phase and amplitude discrepancies as the frequency moves farther beyond 16~Hz. At 17~Hz, eight-step mean flows reduces the error to 0.1554, whereas at 18~Hz its best tested error is 0.2831 at two steps. Diffusion likewise reaches its lowest 18~Hz error at two steps, approximately 0.3770, before becoming less accurate at longer rollouts. 

Frequency extrapolation changes both the background-wavefield condition and the scattered-field target. Increasing frequency alters the wavelength and phase pattern of $U_0$, the frequency-dependent forcing in Equation~\eqref{eq:scattered-helmholtz}, and the interference pattern of $\delta U$. The resulting conditional mapping is therefore less well constrained outside the training band. The selected examples show that mean flows retains an accuracy advantage over diffusion under these shifts, while the growing errors at 17 and 18~Hz identify a limit of the present training range. Extending that range, or introducing frequency-aware physical constraints, is a natural next step for improving extrapolation. 

\subsection{Efficiency Comparison}

The timing experiments use an Intel Core i9-14900KF CPU and an NVIDIA GeForce RTX 5070 GPU with 12~GB of memory. Each configuration uses three warm-up runs followed by ten timed runs, and reported times are medians of the timed runs. The multifrequency benchmark uses ten validation velocity models. The numerical baselines solve frequencies either sequentially with one thread or concurrently with up to 16 worker threads.

\begin{figure}[!tb]
\centering
\includegraphics[width=\columnwidth]{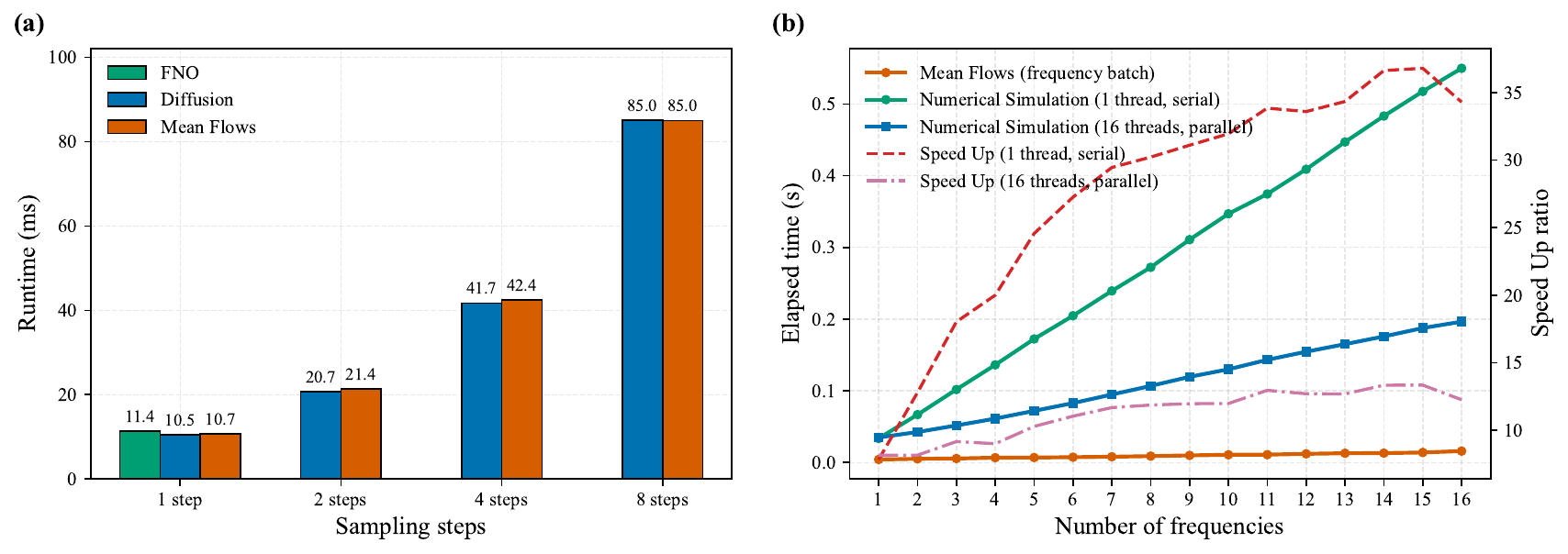}
\caption{Sampling and multifrequency inference efficiency. Panel (a) shows GPU runtimes for ten velocity models at 12~Hz. Panel (b) compares one-step mean flows with sequential single-thread and 16-thread frequency-parallel CPU simulation over 3--18~Hz. Mean flows batches the requested frequencies in one GPU call. The left axis shows seconds per model; the right axis shows numerical runtime divided by mean flows runtime.}
\label{fig:multifrequency-efficiency}
\end{figure}

Fig.~\ref{fig:multifrequency-efficiency}(a) shows that FNO, diffusion, and mean flows have single-pass inference times of the same order, while the costs of diffusion and mean flows increase approximately in proportion to the number of sampling steps. For a batch of ten velocity models at 12~Hz, FNO takes 11.4~ms for its direct prediction, compared with 10.5~ms for one-step diffusion and 10.7~ms for one-step mean flows. These similar single-pass costs are consistent with their comparable parameter budgets: approximately 52.3 million parameters for FNO and 51.6 million for each generative model. Diffusion and mean flows also share the same spatial backbone, giving similar costs at a fixed step count. Across 1, 2, 4, and 8 steps, their runtimes grow to 85.0~ms for both methods at eight steps, as each additional step requires another network evaluation. Where one mean flows update provides sufficient accuracy, it therefore avoids the repeated evaluations required by longer sampling rollouts.

Fig.~\ref{fig:multifrequency-efficiency}(b) compares runtimes as the number of requested frequencies increases from one at 3~Hz to all 16 integer frequencies over 3--18~Hz. For the complete set, mean flows takes 0.0160~s per velocity model, compared with 0.5498~s for sequential single-thread numerical simulation and 0.1965~s for frequency-parallel numerical simulation. These measurements correspond to speed-ups of $34.3\times$ and $12.3\times$, respectively, under Equation~\eqref{eq:speedup}. In the parallel numerical baseline, the number of active worker threads increases with the number of requested frequencies, with one worker per frequency and up to 16 workers for the complete set. This parallelism reduces the numerical runtime by approximately $2.8\times$ relative to sequential execution, although each frequency still requires separate operator assembly, factorization, and solution. Mean flows stacks the frequency-specific background wavefields and velocity conditions along the batch dimension, processing all requested frequencies in a single network call per velocity model. This batching shares inference overhead across frequencies, allowing runtime to grow more slowly than the frequency count.

\section{Discussion}

More sampling steps do not necessarily reduce wavefield error because endpoint accuracy depends on both the sampling approximation and the accuracy of the learned dynamics. For diffusion, finer time discretization can reduce the numerical integration error of the learned reverse process, as illustrated by diffusion ODE solvers such as DPM-Solver \cite{lu2022dpmsolver}, but it does not remove errors in the denoiser itself. A further source of error arises from the states encountered during sampling. Ning et al.~\cite{ning2023input} show that training on noisy reference samples and sampling from previously generated states introduces a distribution mismatch that allows prediction errors to accumulate; their experiments demonstrate that shorter sampling chains can outperform longer ones. In the present wavefield setting, repeated updates can similarly propagate phase and amplitude errors, especially at unseen frequencies where the conditional mapping is less well constrained. For mean flows, the full-interval update predicts the accumulated displacement directly, whereas interval partitioning evaluates successive average velocities at generated intermediate states. Errors in these states and in the interval-specific velocity estimates can offset the benefit of using shorter intervals. This balance is consistent with the lowest mean flows errors occurring at four steps in the CurveVel-A and FlatFault-A comparisons and at two steps for Marmousi at 18~Hz, followed by increases at longer rollouts. The improvements on Marmousi and Overthrust at 16~Hz show the other outcome: additional updates can still refine the prediction when the learned intermediate transport is sufficiently accurate. Sampling length therefore controls how the learned dynamics are followed, rather than guaranteeing a monotonic reduction in error relative to the Helmholtz solution.

Diffusion and Flow Matching models can generate high-quality samples, but many practical tasks require outputs to satisfy constraints beyond resembling the learned data distribution. Existing approaches broadly follow two strategies. The first retrains a generative model to accept the task-specific condition, as in classifier-free guidance \cite{ho2022classifierfree}; the second modifies the sampling process to incorporate additional guidance or data-consistency constraints at each step, as in diffusion posterior sampling, physics-informed diffusion, and physics-guided seismic diffusion \cite{chung2023diffusion,shu2023physics,cheng2026seismic}. The former requires additional training, whereas the latter may rely on assumptions about the forward model or on a particular guidance strategy. D-Flow introduces a third strategy for flow-based generation: it freezes a pretrained generator and optimizes its initial noise by differentiating a task-specific loss through the flow \cite{benhamu2024dflowdt}. The pretrained model then acts as a data-distribution prior, while the differentiable loss expresses the constraint to be satisfied. 

A limitation of the study is its fixed-grid representation: both the conditional input and scattered wavefield are represented on a $70\times70$ grid, restricting direct transfer across resolutions. A change in grid spacing alters the number of samples per wavelength and the sampled geometry of subsurface interfaces, so the learned transport cannot be assumed to retain its accuracy on a different discretization. FunDiff addresses variable discretizations by encoding sampled physical fields into a common latent representation and reconstructing continuous functions at requested spatial coordinates; its training also exposes the encoder to inputs at different sampling densities \cite{fundiff2026}. For scattered-wavefield generation, a corresponding function-space representation could encode the velocity model and analytical background field together with their physical coordinates, learn mean flows transport in a grid-independent latent space, and evaluate the real and imaginary scattered fields at the desired output coordinates. This could allow the same conditional transport to be queried on different grids while retaining the source and frequency information carried by the background field.

\section{Conclusions}

This study develops a conditional mean flows model for frequency-domain acoustic scattered-wavefield generation. The model conditions the transport on the velocity model and the analytical background wavefield, and learns the normalized scattered field through a finite-interval transport path. Trained on OpenFWI over the 3--16~Hz frequency band, the model generalizes to the evaluated out-of-distribution velocity structures and extrapolates to frequencies beyond the training band. In the selected comparisons, one-step mean flows sampling achieves lower relative $L_2$ errors than eight-step diffusion sampling while incurring substantially lower inference cost. In the efficiency experiments, batched generation of 16 frequencies from 3--18~Hz achieves $34.3\times$ and $12.3\times$ speedups over sequential single-thread and 16-thread frequency-parallel CPU numerical simulations, respectively. Together, these results highlight the potential of mean flows for accurate and efficient repeated wavefield simulation in seismic imaging and inversion.

\section*{Acknowledgment}

This research was partly supported by the CNPC Innovation Fund (2024DQ02-0139) and High-performance Computing Platform of Chengdu University of Technology.

\section*{Data and Code Availability}

The code and  data used in this study are publicly available at \url{https://github.com/Lipeng-Lai/MeanFlow4Helmholtz}.

\bibliographystyle{IEEEtran}
\bibliography{example}
\end{document}